# Quantum anomalous Hall effect induced by circularly polarized light on samarium hexaboride surface

*Udai Prakash Tyagi and *Partha Goswami*

*D.B.College, University of Delhi, Kalkaji, New Delhi-110019, India*

*Email of the first author:* uptyagi@yahoo.co.in

**Email of the corresponding author:* physicsgoswami@gmail.com

## Abstract

We examine a time-dependent, surface Hamiltonian for the 3D compound samarium hexaboride based on the slave boson (SB) protocol linked version of the periodic Anderson model reported earlier. The problem of large on-site electron-electron repulsion was reformulated in terms of a holonomic constraint involving a term '$|b|^2$' representing the spatially-independent SB-condensate. In this communication, we show the possible access to the quantum anomalous Hall state due to the normal incidence of circularly polarized light on the surface of the compound in the high frequency limit within the framework of the Floquet theory. The value of '$|b|^2$' is mildly affected by the intensity of incident radiation. The chern number is found to be unity for the right-handed as well as the left-handed circularly polarized light.

## 1.INTRODUCTION

The polarized terahertz radiation provides a potent modus operandi to carry out theoretical proposition and experimental realization, manipulation, and detection of diverse unconventional/novel electronic properties of materials, such as the realization of novel quantum phases like light induced quantum anomalous Hall (QAH) phase **[1-3]**, the topological phase transitions in semi-metals **[4-8]**, the Floquet engineering of magnetism in topological insulator thin films **[9,10]**, and so on. The exotic Floquet topological phases with a high tunability could be realized using this highly efficient and promising platform. In fact, there has been an upsurge on experimental front in the search for topological states (which may be inaccessible in static systems), in solid state **[11]**, ultra cold-atom **[12],** and optical systems **[13]** and so on through the periodic driving of polarized radiation. The circularly polarized radiation field **(CPRF)** is described by a time-periodic (time period $T = 2\pi/\omega$ where ω is the frequency of light) electromagnetic gauge field $\boldsymbol{A}(t)$. Once we have included a gauge field, it is necessary that we make the Peierls substitution which couples lattice electrons to the gauge field. Thus, in the presence of CPRF, the Hamiltonian of a system becomes periodic in time. One can now transform the time-dependent Hamiltonian problem to a time-independent one using the Floquet's theory **[14-19]**. Analogous to the Bloch theory, a solution for the time-dependent Schrodinger equation of the system is obtained involving the Floquet quasi-energy and the Floquet state. The state could be expanded in a Fourier series which makes us arrive at an infinite dimensional eigenvalue equation in the Sambe space **[14].** In the Floquet-Magnus limit **[19]**, the system irradiated by the circularly polarized radiation can be described by an effective, static Hamiltonian.

The minimalistic bulk Hamiltonian of a topological Kondo insulator (TKI) $SmB_6$ is given in refs.**[20-25].** This is a slave boson (SB)protocol related extended version of the periodic Anderson model (PAM) **[22-24]**. Our bulk Hamiltonian in ref. **[23] (**of $SmB_6$) involving $d$ and $f$ fermions (with the corresponding hopping integrals denoted by $t_{d_1}$ and $t_{f_1}$, respectively) captured essential physics of TKI in the presence of the strong coulomb repulsion $U_f$ ( $U_f$ >> $t_{d_1}$) between $f$ electrons on the same site, and the spin-orbit hybridization. The latter is the harbinger of a topological dispensation. In the references **[22-24]**, a detailed exposition of PAM and SB formalism could be found. In the present paper, which is a sequel to that in ref. **[23],** we make use of the Floquet theory **[14-19]** to investigate the system surface interaction with the incident electromagnetic radiation in the high-frequency limit. For this purpose, we consider a low-energy time-independent, surface Hamiltonian ( $H_f$) – a variant of the extended PAM. This Hamiltonian has been obtained by the evanescent wave method. The slab-geometry method could equally well be applied. Interestingly, the radiation field leads to the possibility of the emergence of the quantum anomalous Hall (QAH)state as we find the integer values of the chern number $C$ ( $C = 1$).

The way we structure this paper is as follows: We obtain a low-energy version of the Floquet surface Hamiltonian in section 2. Upon using this Hamiltonian we discuss the possibility of the emergence of a novel phase due to the incident CPRF leading to broken time reversal symmetry (TRS). In section 3, we calculate the eigenvalues and the corresponding eigenvectors of the surface Hamiltonian. We also present the corresponding band spectrum in this section. To gain further insight into the nature of this emergent phase we calculate the chern number in section 4. The communication ends with very brief concluding remarks in section 5.

## 2. MODEL

The periodic Anderson model (PAM) for a topological Kondo insulator $SmB_6$ involving $d$ and $f$ electrons is discussed in refs. **[22-25].** In particular, refs. **[23,24]** present PAM, and its extension using the slave-boson (SB) protocol, in full glory. Upon following these references we write the Hamiltonian matrix in momentum-space as $h(\boldsymbol{K} = (k_x, k_y, k_z), \mu, |b|) = \frac{\widetilde{E_k^d}(\mu,K)}{2}( I^{4\times4} + \gamma^0) + \frac{\widetilde{E_k^f}(\mu,|b|,K)}{2}( I^{4\times4} - \gamma^0) + \vartheta_\mu \gamma^0 \gamma^\mu$ in the basis $(d_{k,\uparrow}\ \ d_{k,\downarrow}\ \ |b|c_{k,\uparrow}\ \ |b|c_{k,\downarrow})^T$ straightway. We represent the creation (annihilation) operators by $d_{k,\tau}^\dagger(d_{k,\tau})$ and $f_{k,\tau}^\dagger = |b|c_{k,\tau}^\dagger\,(|b|c_{k,\tau})$, respectively, for $d$ and $f$ electrons. The index $\tau$ (= ↑, ↓) represents the spin (pseudo-spin) for $d$-($f$-) electrons. The parameter $'b'$ may be complex as the density distribution of the Bose-condensate is represented by a wavefunction with a well-defined amplitude and phase **[23]**. Here-in-after, we shall sometimes denote $|b|$ simply by the letter '$b$'**.** The dispersions $\widetilde{E_k^d}(\mu, K) = -\mu + 3t_{d1} - 3t_{f1} - \mathfrak{b}_d$ and $\widetilde{E_k^f}(\mu, b, K) = -\mu - 3t_{d1} - 3t_{f1} + b^2(\epsilon_f + 6t_{f1}) - \mathfrak{b}_f$, where $(\mathfrak{b}_d, \mathfrak{b}_f)$ are given in the Appendix A. The symbol $I^{4\times4}$ is for the 4× 4 identity matrix, $\mu$ is the chemical potential of the fermion number, and $\vartheta_\mu = (\vartheta_x, \vartheta_y, \vartheta_z)$, where $\vartheta_j = \left(2Vb\sin ak_j\right), j = (x, y, z)$.The anti-commuting($\frac{1}{2}\{\gamma^\mu, \gamma^\nu\}= g^{\mu\nu} I_{4\times4}$) matrices $(\gamma^0, \gamma^1, \gamma^2, \gamma^3, \gamma^5)$ are Dirac matrices in contravariant notations. We note that the $\gamma^0$matrix is hermitian while the $(\gamma^1, \gamma^2, \gamma^3)$matrices are anti-hermitian. The terms $\left(t_{d1}, t_{f1}\right)$, $(t_{d2}, t_{f2})$, and $\left(t_{d3}, t_{f3}\right)$, respectively, are the *NN*, *NNN*, and *NNNN* hopping parameters for $d$ and $f$ electrons; $\epsilon_f$ is the onsite energy of the $f$ electrons and $a$ = 0.413 nm is the crystal structure lattice constant of $SmB_6$ bulk. The system shows the bulk metallic as well as the bulk insulating phases are determined by the sign of $t_{f1}$. It is positive for the former and negative for the latter phase **[22-25]**. The negative sign of $t_{f1}$ is also necessary for the band inversion, which induces the topological state **[22-25]**. Throughout the paper, we choose $t_{d_1}$to be the unit of energy. Furthermore, it is straightforward to show that, without spin-orbit interactions, the Anderson Model is both time-reversal symmetric and inversion symmetric.

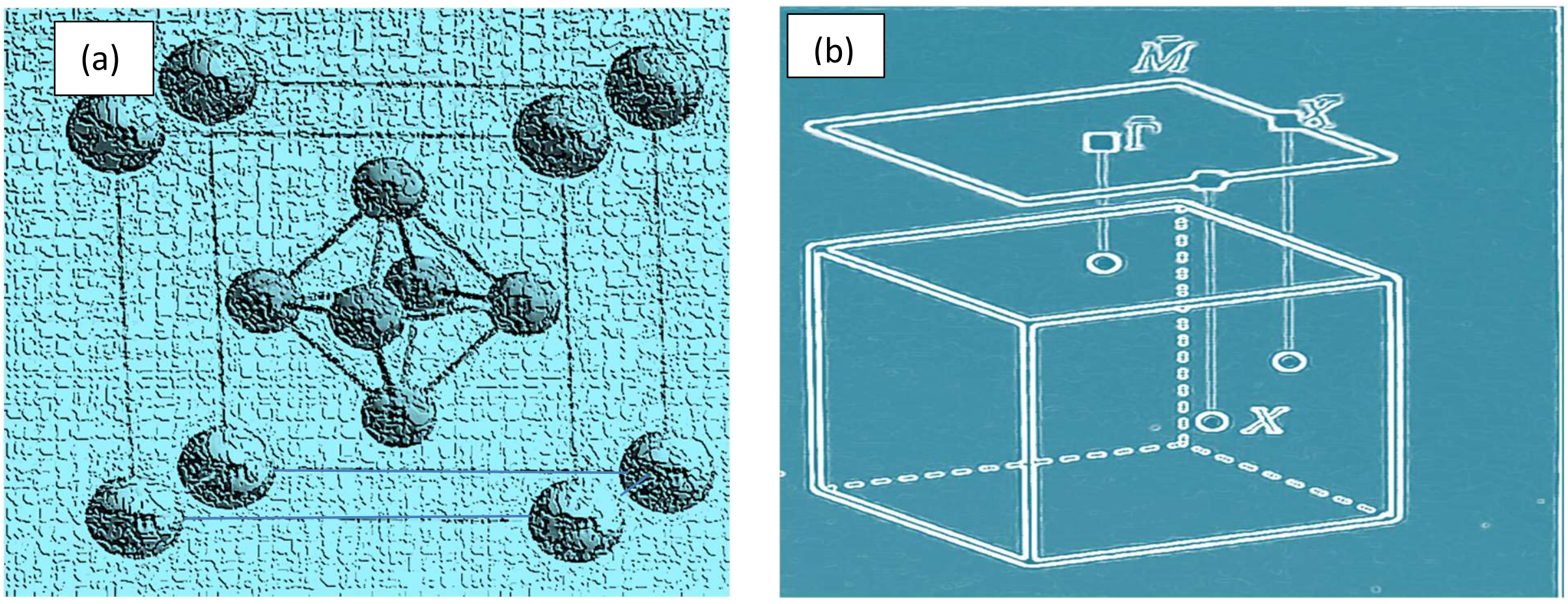

**Figure 1. (a)** A diagrammatic representation of $SmB_6$ crystal structure (with cubic lattice constant $a$ = 0.413 nm). The Sm ions are located at the corners and the $B_6$ octahedron at the center of the cubic lattice. **(b)**The high symmetry points of the bulk Brillouin zone (BZ) are $\Gamma(0,0,0)$, $X\{(\pi,0,0), (0,\pi,0), (0,0,\pi)\}$, $M\{(\pi,\pi,0), (\pi,0,\pi), (0,\pi,\pi)\}$ and $R(\pi,\pi,\pi)$. The projection of the $X$ points in the (001) surface BZ gives the points $\bar{\Gamma}$ (0,0) and $\bar{X}$ $(\pi, 0)$.

In order to obtain surface Hamiltonian $H\left(k_x, k_y, \mu, b\right)$, earlier **[22,23]** we have adopted the method involving the slab-geometry considering a set of orthonormal basis states. For the sake brevity and simplicity we lean on the heuristic evanescent wave approach (EWA) **[23,26]** here. The details are given in the Appendix A**.** We treat the model Hamiltonian matrix in (1) in the low-energy limit near $\bar{\Gamma}$ (0,0) point. As shown in the appendix A, the low-energy surface (2D) Hamiltonian can be conveniently written as $H\left(k_x, k_y, \mu, b, d\right) = \frac{\epsilon(k,d,\mu,b)+\vartheta(k,b,d)}{2}(I^{4\times4} + \gamma^0) + \frac{\epsilon(k,d,\mu,b)-\vartheta(k,b,d)}{2}(I^{4\times4} - \gamma^0) + iA_1ak_x\gamma^2 + A_1ak_y\gamma^0\gamma^2 - iA_1(\frac{a}{d})\gamma^3$, where $A_1 = 2Vb$. In this limit, the functions $\epsilon(k,\mu,d,b) = \frac{(\widetilde{E_k^d}(\mu,k)+\widetilde{E_k^f}(\mu,b,k))}{2}$ and $\vartheta(k,\mu,d\,,b) = \frac{(\widetilde{E_k^d}(\mu,k)-\widetilde{E_k^f}(\mu,b,k))}{2}$ have been approximated to $O(a^2k^2)$. These functions are given in Appendix A.

We now assume the normal incidence of CPRF on the surface $SmB_6$. Suppose the angular frequency of the radiation field incident on the surface is $\omega \sim 10^{15} radian - s^{-1}$, the period $T = 2\pi/\omega \sim 10^{-14}s$, and wavelength $\lambda_{in} \approx 1500\ nm$. The radiation field $\boldsymbol{E}(t)$ may be expressed in terms of the electric scalar potential (assumed to be zero) and the time-varying magnetic vector potential $\boldsymbol{A}(t) = \boldsymbol{A}(t+T) = \boldsymbol{A_0}(sin(\omega t), sin(\omega t + \psi), 0)$ through the relation: $\boldsymbol{E}(t) = -\frac{\partial \boldsymbol{A}(t)}{\partial t} = -\boldsymbol{E_0}(cos(\omega t), cos(\omega t + \psi), 0), \boldsymbol{E_0} = \boldsymbol{A_0}\omega$. In particular, when the phase $\psi = 0\ or\ \pi$, the radiation field is linearly polarized. When $\psi = \frac{\pi}{2}\left(\psi = -\frac{\pi}{2}\right)$, the radiation field is left-handed (right-handed) circularly polarized. Upon taking the electromagnetic field into consideration the Hamiltonian above becomes time dependent. Once we have included a gauge field, it is necessary that we make the Peierls substitution: $H\left(\boldsymbol{k} - \frac{e}{\hbar}\boldsymbol{A}(t), \mu, b\right) = H_f(\boldsymbol{k}, t, b)$, where $\boldsymbol{k} = (k_x, k_y)$. The quantity $I = (aA_0)^2 = \left(\frac{aeE}{\hbar\omega}\right)^2$ corresponds to the dimensionless radiation intensity. Now the Floquet theory **[14-19]** can be applied to our time-periodic Hamiltonian $H_f(t) = H_f(t+T)$ to obtain a static effective Hamiltonian $H_f(k, d, aA_0, \mu, b)$. The details have been presented in the Appendix B**.** We have shown that, only when the incident radiation field is left-(or, right-) handed circularly polarized, the time reversal symmetry (TRS) is broken by $H_f(k, q, aA_0, b)$.Another possible way of visualizing TRS breaking will be to write the Hamiltonian $H_f(k, q, aA_0, b)$ in the basis $(d_{k,\uparrow}\ \ d_{k,\downarrow}\ \ bc_{k,\uparrow}\ \ bc_{k,\downarrow})^T$. We obtain

$$H_f = \epsilon_{OP}\, I^{4\times4} + \frac{\vartheta_{OP}^{+}+\vartheta_{OP}^{-}}{2}\gamma^0 + 2C_\mu\Sigma^\mu + A_{1OP}^{+}ak_y\frac{\gamma^0\,\gamma^2-i\gamma^1}{2} + A_{1OP}^{-}ak_y\frac{\gamma^0\,\gamma^2+i\gamma^1}{2}$$

$$+A_{1OP}^{+}ak_x\frac{\gamma^0\,\gamma^1+i\gamma^2}{2} - A_{1OP}^{-}ak_x\frac{\gamma^0\,\gamma^1-i\gamma^2}{2} + \frac{i(A_{1OP}^{-}-A_{1OP}^{+})aq}{2}\gamma^0\gamma^5 - \frac{i(A_{1OP}^{-}+A_{1OP}^{+})aq}{2}\gamma^3, \quad (1)$$

where $C_\mu = \left(0, 0, M = \frac{\vartheta_{OP}^{+}-\vartheta_{OP}^{-}}{2}\right), \Sigma^\mu = \left(\frac{1}{2}\right)\epsilon^{\mu\nu\rho}\sigma_{\nu\rho}, q = d^{-1}$, and $\sigma_{\nu\rho} = \left(\frac{i}{2}\right)[\gamma_\nu, \gamma_\rho]$. Given the Dirac matrices obeying the anticommutation relations mentioned above, we may define the spin matrices as $\frac{1}{2}\sigma^{\nu\rho}$. It may be noted that these matrices obey the same commutation relations as the generators $J^{\mu\nu}$of the continuous Lorentz group. Moreover, the commutation relations of $J^{\mu\nu}$ with a Lorentz vector are similar to the commutators $[\frac{1}{2}\sigma^{\nu\rho}, \gamma^\mu]$. As a matter of fact, all such four-band models could be written out in terms of the Dirac matrices. This 2D Hamiltonian is presented in the alternative basis $(d_{k,\uparrow}\ \ bc_{k,\downarrow}\ \ d_{k,\downarrow}\ \ bc_{k,\uparrow})^T$ in the appendix A. Equation (1) or the one in the alternative basis in the appendix A is our periodically driven model Hamiltonian to be used in our analysis. The functions $\epsilon_{OP} = \epsilon_{OP}\,(k, q\,, \mu, b), A_{1OP}^{(\alpha)}(b, aA_0, \psi)$ and $\vartheta_{OP}^{(\alpha)} = \vartheta_{OP}^{(\alpha)}(k, q, b, aA_0, \psi)$ (the superscript $\alpha$ stands for $\pm$) are defined below:

$$\epsilon_{OP}(k, q\,, \mu, b) = \epsilon_0(\mu, b) - D_1(b)a^2q^2 + D_2(b)a^2k^2 + a^2A_0^2D_2(b) + O(a^4k^4), \quad (2)$$

$$\vartheta_{OP}^{(\alpha)}(k,q,b)=\vartheta_0(b)-B_1(b)a^2q^2+B_2(b)a^2k^2-\left(a^2A_0^2B_2+\alpha\left(\frac{a^2A_0^2}{\hbar\omega}\right)sin\psi\ A_1^2\right), \quad (3)$$

$$A_{1OP}^{(\alpha)}=A_1\left(1+2\alpha B_2 sin\psi\left(\frac{a^2A_0^2}{\hbar\omega}\right)\right),\ A_1=\ 2Vb,\ C_3=M=[-2\left(\frac{a^2A_0^2}{\hbar\omega}\right)sin\psi\ A_1^2]\ . \quad (4)$$

Along with the off-diagonal terms, now the diagonal (mass) terms $[\frac{\vartheta_{OP}^{+}+\vartheta_{OP}^{-}}{2}\gamma^0+2C_\mu\Sigma^\mu]$ ,dependent on the intensity and the polarization of radiation, appear in the Hamiltonian which implies that the system is now without the chiral symmetry. Moreover, we notice from Eq.(2)-(4) (and the appendix A) that CPRF not only renormalizes terms involving *d* and *f* electron hopping integrals but also does the renormalization of the hybridization parameter(HP).The term $C_3=M$ acts as the pseudo-magnetic exchange energy responsible for breaking the time reversal invariance (TRS).

The term $C_3=M$ in (1) depends on the intensity and frequency of the incident radiation and the hybridization parameter. The modulation of this emergent pseudo-magnetization is possible by circularly polarized radiation field (CPRF): it increases with intensity of the incident radiation and decreases with the frequency. Since $\psi=\frac{\pi}{2}\left(\psi=-\frac{\pi}{2}\right)$, corresponds to the left-handed (right-handed) CPRF, we have $M<0$ for the former and $M>0$ for the latter. The latter case corresponds to the parallel spin configuration (pseudo-ferromagnetism) while the former to the anti-parallel one(pseudo-antiferromagnetism). Thus, as TRS is broken due to the incidence of the CPRF on the $SmB_6$ surface, the quantum anomalous Hall (QAH) insulator state is likely to be accessible provided we show that the Chern number ($C$), associated with anomalous Hall conductivity (AHC), has integer values. In the next section, we calculate the energy eigenvalues and the corresponding eigenvectors of the matrix (1) resulting from the interaction with radiation. We calculate the Chern number $C$ in the section 4.

## 3. SURFACE STATE

The single-particle excitation spectrum (SPES)and the spectral gap, representing allowed energy values for electron in a system, are central feature to investigate as they provide important insights into the understanding the system. The eigenvalues ($E_\alpha$) of the matrix (1), given by the quartic below, gives us SPES of the system driven periodically by the polarized radiation:

$$E_\alpha^4+\gamma_{1OP}(k,b)\ E_\alpha^3+\gamma_{2OP}(k,b)\ \ E_\alpha^2\ +\gamma_{3OP}(k,b)\ E_\alpha+\gamma_{4OP}(k,b\ )=0\ (\alpha=1,2,3,4)\ (6)$$

where the coefficients $\gamma_{\alpha OP}\ (k,b)$ ($\alpha$ = 1,2,3,4) of the quartic are given by

$$\gamma_{1OP}=-\sum_\alpha \varepsilon_\alpha, \gamma_{2OP}=\sum_{\mu\neq\nu}\varepsilon_\mu\varepsilon_\nu-(A_{1O}^{+}+A_{1OP}^{-})(aq)^2-({A_{1OP}^{+}}^2+{A_{1OP}^{-}}^2)(ak)^2,\ (7)$$

$$\gamma_{3OP}(k,b)=-(\varepsilon_1+\varepsilon_2)\varepsilon_3\varepsilon_4-(\varepsilon_3+\varepsilon_4)\varepsilon_1\varepsilon_2+(\varepsilon_1+\varepsilon_2){A_{1OP}^{-}}^2(ak)^2+(\varepsilon_3+\varepsilon_4){A_{1OP}^{+}}^2(ak)^2$$
$$+(\varepsilon_1+\varepsilon_4)\ {A_{1O}^{-}}^2(aq)^2+(\varepsilon_2+\varepsilon_3){A_{1OP}^{+}}^2(aq)^2,\ (8)$$

$$\gamma_{4OP}(k)=\prod_\mu\varepsilon_\mu-{A_{1OP}^{-}}^2(ak)^2(\varepsilon_1\varepsilon_2)-{A_{1OP}^{+}}^2(ak)^2(\varepsilon_3\varepsilon_4)+{A_{1OP}^{-}}^2{A_{1OP}^{+}}^2(ak)^4+{A_{1OP}^{-}}^2{A_{1OP}^{+}}^2(aq)^4$$

$$-A_{1OP}^{-}{}^{2}(aq)^2(\varepsilon_1\varepsilon_4) - A_{1OP}^{+}{}^{2}(aq)^2(\varepsilon_2\varepsilon_3) - 2A_{1OP}^{-}{}^{2}A_{1OP}^{+}{}^{2}(aq)^2((ak_x)^2 - (ak_y)^2), \quad (9)$$

and

$$\varepsilon_1 = \epsilon_{OP} + \vartheta_{OP}^{+}, \varepsilon_2 = \epsilon_{OP} - \vartheta_{OP}^{+}, \varepsilon_3 = \epsilon_{OP} + \vartheta_{OP}^{-},\ \varepsilon_4 = \epsilon_{OP} - \vartheta_{OP}^{-}. \quad (10)$$

The functions $\epsilon_{OP} = \epsilon_{OP}\ (k, q\ , \mu, b), A_{1OP}^{\pm}$, and $\vartheta_{OP}^{\pm} = \vartheta_{OP}^{\pm}(k, q, b)$ are defined by Eqs. (2)-(4). In view of the Ferrari's solution, we find the roots of the quartic in (6) as

$$E_\alpha(k, l = \pm 1, s = \pm 1, q, b, I) = s\sqrt{\frac{\eta_{OP}(k)}{2} - \frac{\gamma_{1OP}(k,b)}{4} + l\left(b_{OP}(k) - \left(\frac{\eta_{OP}(k)}{2}\right) + s\, c_{OP}(k)\sqrt{\frac{2}{\eta_{OP}(k)}}\right)^{\frac{1}{2}}}. \quad (11)$$

where α = 1,2,3,4, $s = \pm 1$ is the spin index and $l = \pm 1$ is the band-index. The spin- up(down) ($s = \pm 1$) conduction band ( $l = +1$), and the spin-up (down) ($s = \pm 1$) valence bands ( $l = -1$), are denoted, respectively, by $E_1(l = +1, s = +1, k, b, I)$, $E_2(l = +1, s = -1, k, b, I)$, $E_3(l = -1, s = +1, k, b, I)$, and $E_4(l = -1, s = -1, k, b, I)$. The functions appearing in Eq. (11) are given by

$$\eta_{OP}(k) = \frac{2b_{OP}(k)}{3} + \left(\Delta_{OP}(k) - \Delta_{0O}\ \ (k)\right)^{\frac{1}{3}} - \left(\Delta_{OP}(k) + \Delta_{0OP}(k)\right)^{\frac{1}{3}}, \quad (12)$$

$$\Delta_{0OP}(k) = \left(\frac{b_{OP}^3(k)}{27} - \frac{b_{OP}(k)d_{OP}(k)}{3} - c_{OP}^2(k)\right),$$

$$\Delta_{OP}(k) = \left(\frac{2}{729}b_{OP}^6 + \frac{4d_{OP}^2 b_{OP}^2}{27} + c_{OP}^4 - \frac{d_{OP}b_{OP}^4}{81} - \frac{2b_{OP}^3}{27} + \frac{2c_{OP}^2 b_{OP} d_{OP}}{3} + \frac{d_{OP}^3}{27}\right)^{1/2}, \quad (13)$$

$$b_{OP}(k) = \left\{\frac{3\gamma_{1OP}(k,b)^2 - 8\gamma_{2OP}(k,b)}{16}\right\},\ c_{OP}(k) = \left\{\frac{-\gamma_{1OP}(k,b)^3 + 4\gamma_{1OP}(k,b)\gamma_{2OP}(k,b) - 8\gamma_{3OP}(k,b)}{32}\right\}, \quad (14)$$

$$d_{OP}(k) = \frac{-3\gamma_{1OP}(k,b)^4 + 256\gamma_{4OP}(k,b) - 64\gamma_{1OP}(k,b)\gamma_{3O}\ \ (k,b) + 16\,\gamma_{1OP}(k,b)^2\gamma_{2OP}(k,b)}{256}. \quad (15)$$

As noted in section 2, the quantity $I = (aA_0)^2 = \left(\frac{aeE}{\hbar\omega}\right)^2$ corresponds to the dimensionless radiation intensity. In order to obtain the eigenvalues of Eq. (1) when the intensity of the incident radiation is equal to zero, we simply need to put $I = (aA_0)^2 = 0$ in Eq. (11). We have plotted these eigenvalues in ref.**[23]** as a function of the dimensionless wavenumber. The values of the crucial non-free parameter $|b|$ was determined by the minimization of thermodynamic potential per unit volume (see ref.**[22]**). The graphical representations of the four-band spectrum showed the Dirac point feature at $\boldsymbol{k} = (0,0)$, $(\pm 1,0)$, and $(0, \pm 1)$. This is in agreement with the seminal work of Lu et al.**[27]**. On a quick side note, it has been possible to present the Hamiltonian (1) in the block-diagonal form in a certain case as discussed in the appendix A (Eq. (A.8)). We shall focus presently on the non-block diagonal form given by (1).

The graphical representations in Figure 2 correspond to the energy eigenvalues $E_\alpha(s,\boldsymbol{k},b,I)$ plotted as function of the wave number. These plots are obtained using the low energy Hamiltonian in our calculation around the $\overline{\Gamma}$ point. A band gap is the central requirement for QAH to exist. If the gap does not exist, calculating the associated topological invariant called the 'Chern number' (which is basically the Berry curvature (BC) flux integral in momentum space) becomes infructuous. We observe a band-gap at the $\overline{\Gamma}$ point between the bands closer to the chemical potential $\mu = 0$ as in Figures 2(a) (for the left-handed CPRF) and 2(b) (for the right-handed CPRF) with $aA_0 = 1.3678$. Here, the chemical potential lies within the band gap, i.e., the system is in the insulating state. Furthermore, since $E_\alpha(k, l = \pm1, s = \pm1, q, b, I) \neq E_\alpha(-k, l = \pm1, s = \mp1, q, b, I)$, the time reversal symmetry (TRS) are not preserved by Eq.(1). As discussed in appendix B, we arrive at the same conclusion through a different approach. This leads to non-zero BC. All these points show that the investigation based on the model Hamiltonian (1) presented here should have access to the integer values of the chern number ($C$).

In Figure 2(c) and 2(d)we have once again the plots of the energy eigenvalues as function of the wave number for $aA_0 = 0.35$ for the left-handed CPRF and the right-handed CPRF, respectively. Here, though TRS is broken ($aA_0 \neq 0$), we do not expect $C$ to have integer value. The reason being, apart from the valence /the conduction band near degeneracy, the appearance of gap closing and the partially filled conduction band are in place in these cases. Thus, for the lower intensity of the incident radiation, the surface state is conducting. Also, the band gap, required for QAH state to exist, appears when the intensity of the radiation is at a higher value. The Figures 2(e) and 2(f) represent the two-dimensional plots of the surface band spectrum for the plane polarized radiation. While the lower intensity plot (Figure 2(e)) is similar to that in 2(c), the figure 2(f) displays avoided crossing and corresponds to the higher intensity. The other parameter values used are $t_{d_1} = 1$, $t_{f_1} = -0.53$, $t_{d_2} = 0.01$, $t_{f_2} = 0.01$, $t_{d_3} = 0.001$, $t_{f_3} = 0.001$, $\epsilon_f = -0.02$, $b = 0.8426/0.8123/0.8132$ (see Appendix C for an outline of the approach to obtain numerical value of '$b$'), $q = 0.10$, and $\mu = 0$. Our graphical representations lead to the fact that, due to the light-matter interaction induced pseudo-magnetism, the emergent unconventional phase corresponds to a novel state with gap opening or avoided crossing depending on the intensity of the incident radiation.

The surface states linked to the energy eigenvalues in (11) could be written as $|u^{(\alpha)}(\mathrm{k})\rangle = N_\alpha^{-\frac{1}{2}}\phi_\alpha(k)$, where $N_\alpha = [\,\psi_1^{(\alpha)^*}(k)\psi_1^{(\alpha)}(k) + \psi_2^{(\alpha)^*}(k)\psi_2^{(\alpha)}(k) + \psi_3^{(\alpha)^*}(k)\psi_3^{(\alpha)}(k) + \psi_4^{(\alpha)^*}(k)\psi_4^{(\alpha)}(k)]$, and $\phi_\alpha(k)$ is the transpose of the row vector $(\,\psi_1^{(\alpha)}(k) \quad \psi_2^{(\alpha)}(k) \quad \psi_3^{(\alpha)}(k) \quad \psi_4^{(\alpha)}(k))$, $\alpha = 1,2,3,4$. The elements $\psi_j^{(\alpha)}(k)$ are given by $\psi_1^{(\alpha)}(k) = \Upsilon_1^{(\alpha)} + i\Upsilon_2^{(\alpha)}$, $\psi_2^{\alpha}(k) = \varkappa_1^{(\alpha)} + i\varkappa_2^{(\alpha)}$, $\psi_3^{\alpha}(k) = \Delta_1^{(\alpha)} + i\Delta_2^{(\alpha)}$, and $\psi_4^{(\alpha)}(k) = \omega^{(\alpha)} + i0$, where for the $\alpha^{\mathrm{th}}$ band

$$\Upsilon_1^{(\alpha)} = \{-A_{1OP}^{+}{A_{1OP}^{-}}^2((aq)^2(ak_x) + (ak)^2(ak_x)) + A_{1OP}^{+}(ak_x)(E_\alpha - \varepsilon_3)(E_\alpha - \varepsilon_4)\}, \quad (16)$$

$$\Upsilon_2^{(\alpha)} = \{-A_{1OP}^{+}{A_{1OP}^{-}}^2((aq)^2(ak_y) - (ak)^2(ak_y)) - A_{1OP}^{+}(ak_y)(E_\alpha - \varepsilon_3)(E_\alpha - \varepsilon_4)\}, \quad (17)$$

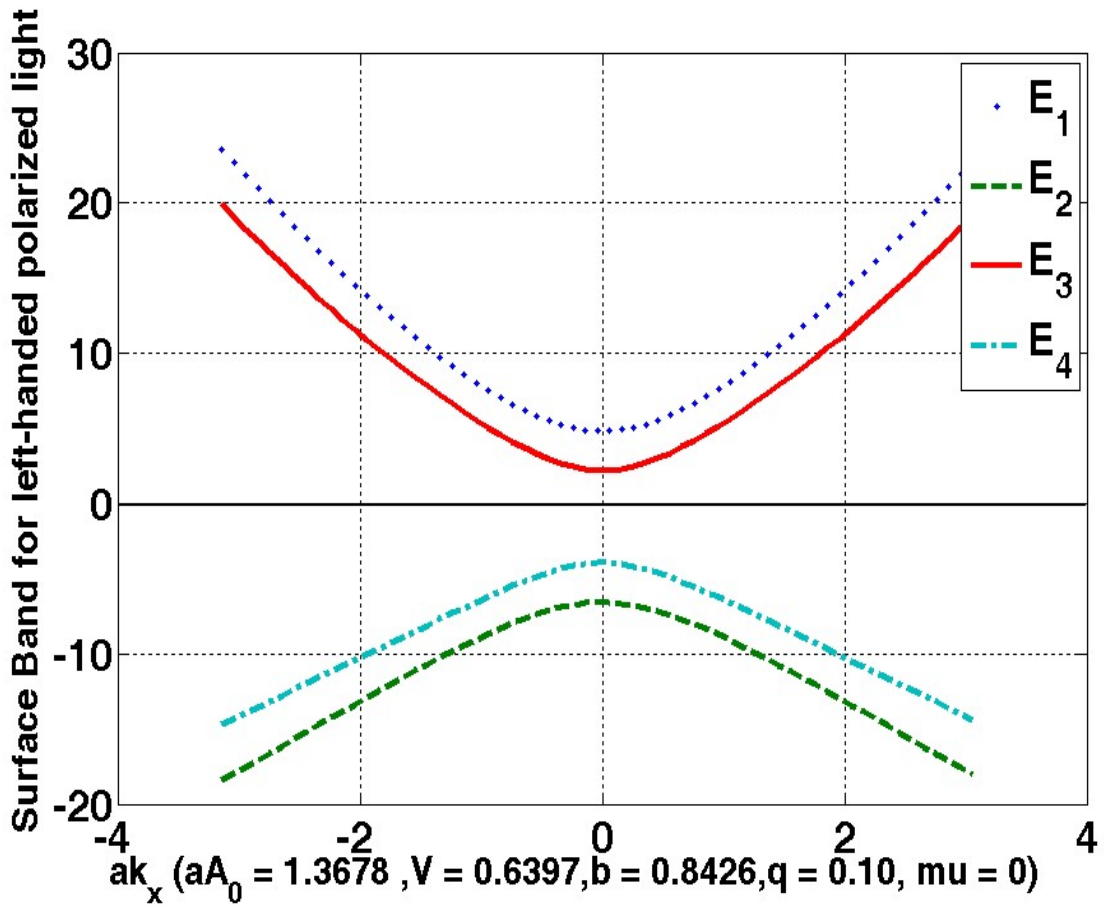
Surface Band for left-handed polarized light
$ak_x$ ($aA_0$ = 1.3678 ,V = 0.6397,b = 0.8426,q = 0.10, mu = 0)
$E_1$
$E_2$
$E_3$
$E_4$

(a)

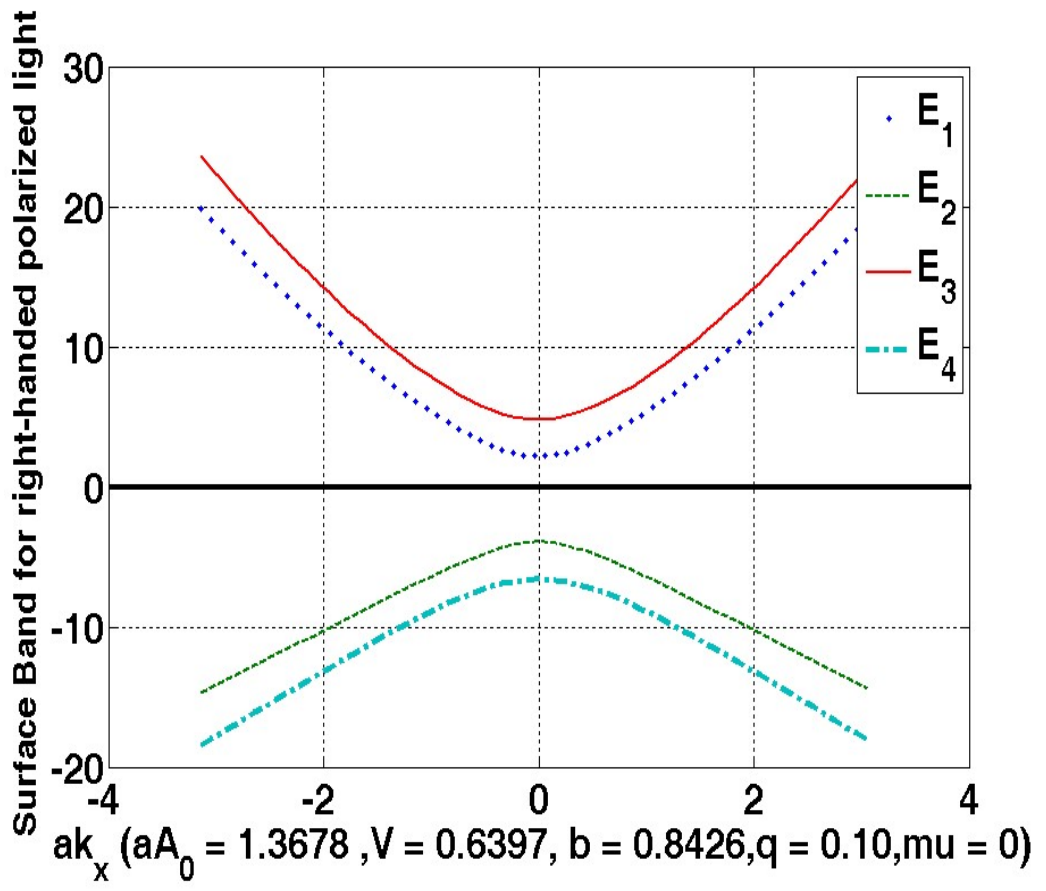
Surface Band for right-handed polarized light
$ak_x$ ($aA_0$ = 1.3678 ,V = 0.6397, b = 0.8426,q = 0.10,mu = 0)
$E_1$
$E_2$
$E_3$
$E_4$

(b)

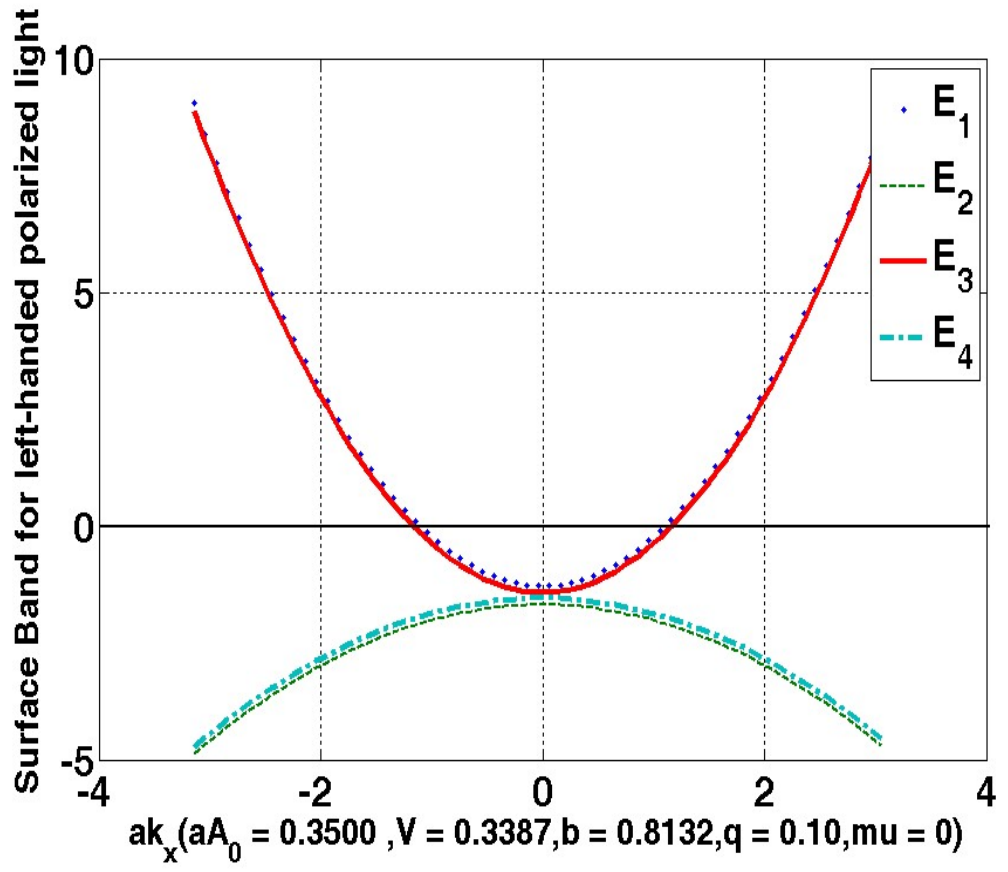
Surface Band for left-handed polarized light
$ak_x$($aA_0$ = 0.3500 ,V = 0.3387,b = 0.8132,q = 0.10,mu = 0)
$E_1$
$E_2$
$E_3$
$E_4$

(c)

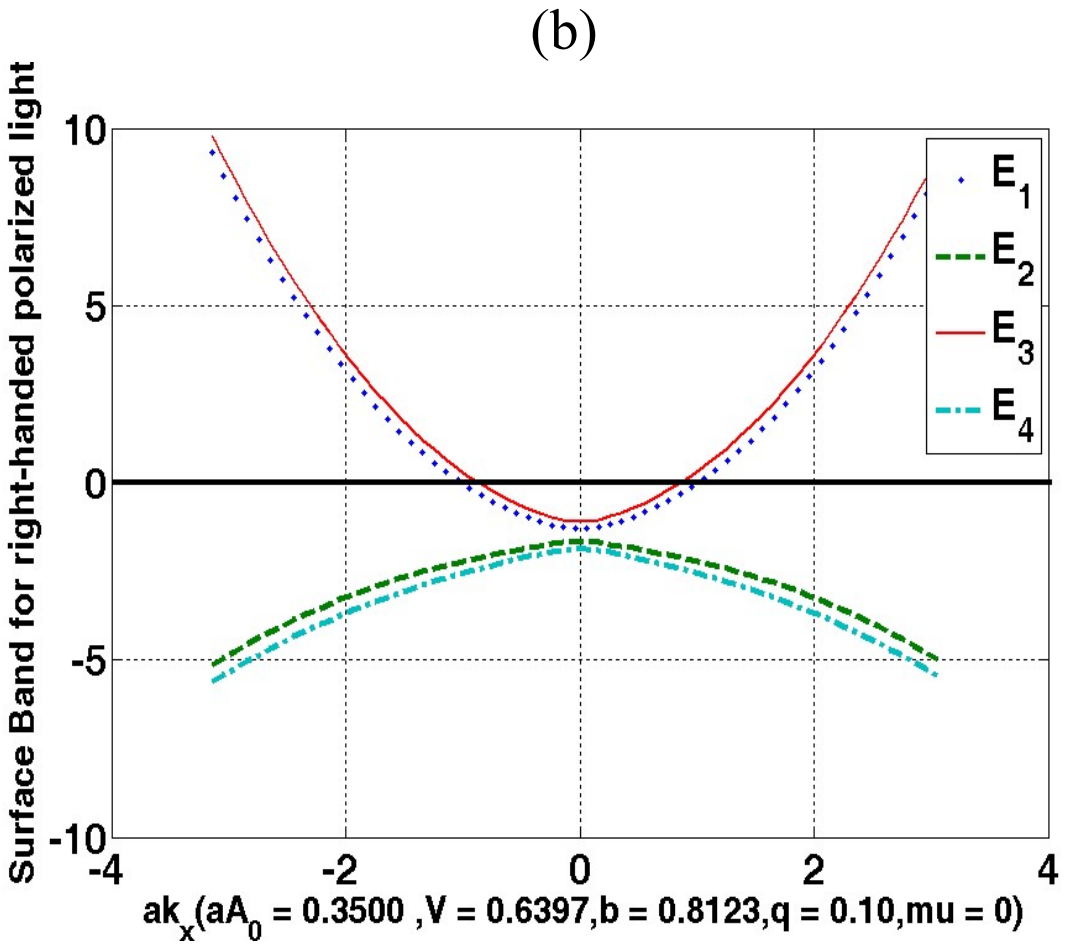
Surface Band for right-handed polarized light
$ak_x$($aA_0$ = 0.3500 ,V = 0.6397,b = 0.8123,q = 0.10,mu = 0)
$E_1$
$E_2$
$E_3$
$E_4$

(d)

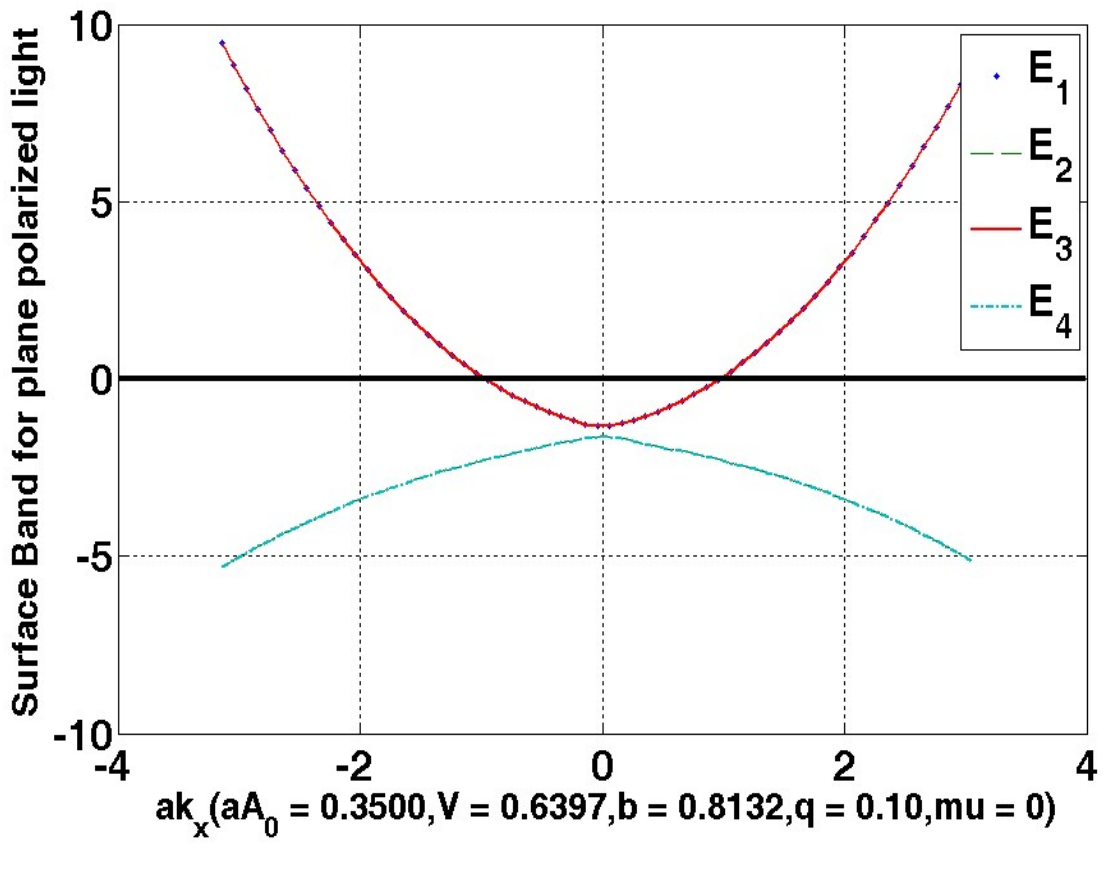
Surface Band for plane polarized light
$ak_x$($aA_0$ = 0.3500,V = 0.6397,b = 0.8132,q = 0.10,mu = 0)
$E_1$
$E_2$
$E_3$
$E_4$

(e)

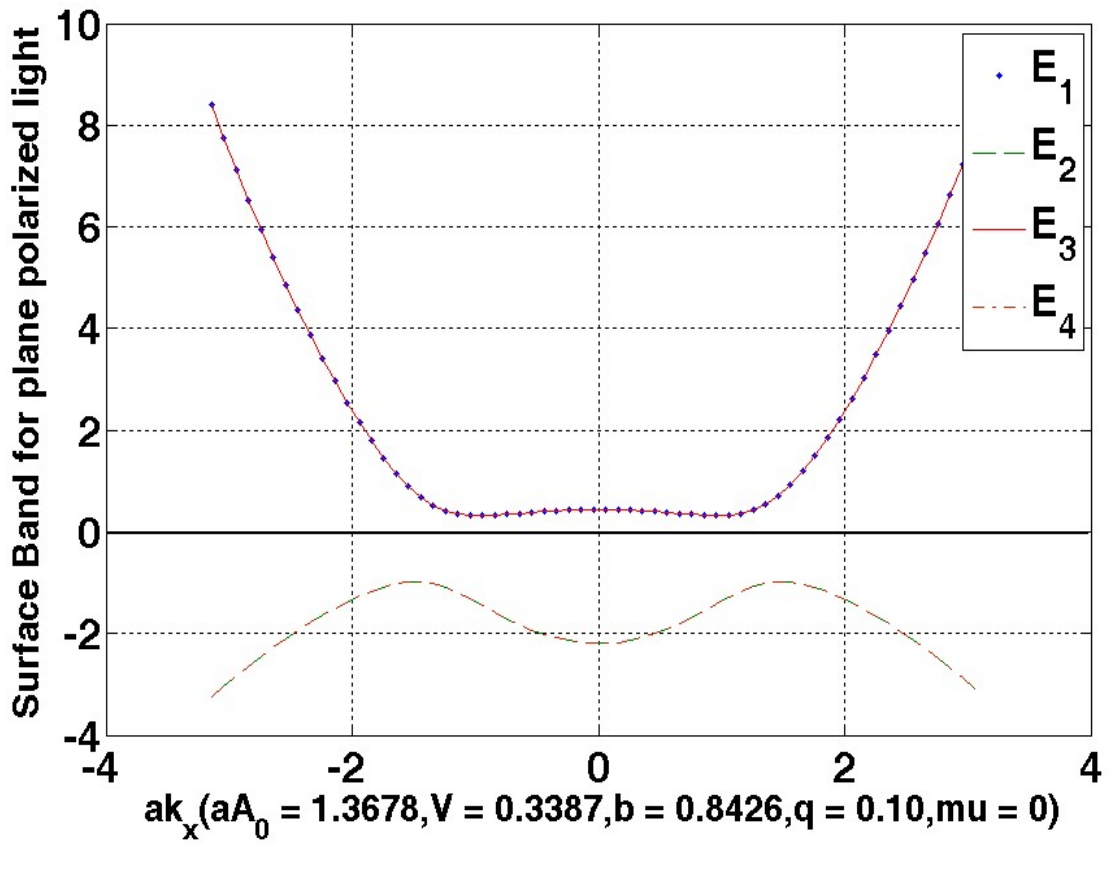
Surface Band for plane polarized light
$ak_x$($aA_0$ = 1.3678,V = 0.3387,b = 0.8426,q = 0.10,mu = 0)
$E_1$
$E_2$
$E_3$
$E_4$

(f)

**Figure 2.** The 2D plots of the single-particle excitation spectrum as a function of the wavenumber component $k_x$ ( $k_y = 0, k_z = 0$).The numerical values of the parameters used in the plots are $t_{d_1} = 1$, $t_{f_1} = -0.53$, $t_{d_2} = 0.01$, $t_{f_2} = 0.01$, $t_{d_3} = 0.001$, $t_{f_3} = 0.001$, $q = 0.10$, $\epsilon_f = -0.02$, and $\mu = 0$. ($a$) and (b) $V = 0.6397$, $b = 0.8426$, and $aA_0 = 1.3678$. ($c$) $V = 0.3387$, $b = 0.8132$, and $aA_0 = 0.3500$.(d) $V = 0.6397$, $b = 0.8123$, and $aA_0 = 0.3500$. (e) $V = 0.6397$, $b = 0.8132$, and $aA_0 = 0.3500$.(f) $V = 0.3387$, $b = 0.8426$, and $aA_0 = 1.3678$.The figures (a) and (c) correspond to left-handed circularly polarized light, while the figures (b) and (d) correspond to right-handed circularly polarized light. The figures (e) and (f) correspond to plane polarized light. The solid horizontal line represents the Fermi energy.

$$\varkappa_1^{(\alpha)} = \left\{-2A_{1OP}^{-}{A_{1OP}^{+}}^2(aq)(ak_x)(ak_y)\right\}, \tag{18}$$

$$\varkappa_2^{(\alpha)} = \left\{ A_{1OP}^{-}(aq)(E_\alpha - \varepsilon_1)(E_\alpha - \varepsilon_4) - A_{1OP}^{-}{A_{1OP}^{+}}^2(aq)^3 - A_{1OP}^{-}{A_{1OP}^{+}}^2(aq)(a^2k_x^2 - a^2k_y^2)\right\}, \tag{19}$$

$$\varDelta_1^{(\alpha)} = \{{A_{1OP}^{-}}^2(E_\alpha - \varepsilon_1) + {A_{1OP}^{+}}^2(E_\alpha - \varepsilon_3)\}(aq)(ak_y), \tag{20}$$

$$\varDelta_2^{(\alpha)} = \{-{A_{1OP}^{-}}^2(E_\alpha - \varepsilon_1) + {A_{1OP}^{+}}^2(E_\alpha - \varepsilon_3)\}(aq)(ak_x), \tag{21}$$

$$\mathbb{O}^{(\alpha)} = (E_\alpha - \varepsilon_1)(E_\alpha - \varepsilon_3)(E_\alpha - \varepsilon_4) - {A_{1OP}^{-}}^2(a^2k_x^2 + a^2k_y^2)(E_\alpha - \varepsilon_1) - {A_{1OP}^{+}}^2(aq)^2(E_\alpha - \varepsilon_3). \tag{22}$$

Here $N_\alpha = {\Upsilon_1^{(\alpha)}}^2 + {\Upsilon_2^{(\alpha)}}^2 + {\varkappa_1^{(\alpha)}}^2 + {\varkappa_2^{(\alpha)}}^2 + {\varDelta_1^{(\alpha)}}^2 + {\varDelta_2^{(\alpha)}}^2 + {\mathbb{O}^{(\alpha)}}^2$. The calculation of eigenstate is the first step to obtain BC. In the next section we calculate the Chern number ($C$) linked to the Berry phase concept and provide evidence that the emergent novel state is could be the quantum anomalous Hall (QAH) phase.

## 4.Quantum anomalous Hall effect

We have seen above that, through the periodic driving of polarized radiation, we are able unlock a new route towards the possible engineering of the quantum anomalous Hall effect (QAHE) due to the presence of the term $C_3 = M\left[-2\left(\frac{a^2A_0^2}{\hbar\omega}\right) sin\psi\ A_1^2\right] \neq 0$. The term acts as the pseudo-magnetic exchange integral. To bolster the possibility of AHE, we calculate the Berry curvature (BC) and the Chern number below.

In the absence of external magnetic fields, the phenomenon of QAHE is largely governed by BC. A transverse velocity to the carrier electrons contributes to QAHE. We consider our case of the four-level system, the Hamiltonian of the system is given by (1) and the normalized eigenstates $|u^{(\alpha)}(\mathrm{k})\rangle$ given by (16) to calculate the anomalous Hall conductivity (AHC). The expression of AHC is $\sigma_{AH} = (\frac{e^2}{h}) \sum_{\alpha\in \text{occupie}} \int_{BZ} \frac{d^2k}{(2\pi)^3} f(E_\alpha(k)-\mu)\, \Omega_\alpha^z(\mathrm{k})$, where μ is the chemical potential of the fermion number, α is the occupied band index, $f(E_\alpha(k)-\mu)$ is the Fermi- Dirac distribution function and $\Omega_\alpha^z(k)$ is the $z$-component of the Berry curvature (BC) for the $\alpha^{th}$ occupied band.To obtain AHC, we calculate first BC using the Kubo formula **[28,29]**

$$\Omega_\alpha^z(k) = -2\left[Im \sum_{\beta\neq\alpha}(E_\alpha(k) - E_\beta(k))^{-2} \left\langle u^{(\alpha)}(\mathrm{k})\middle|\frac{\partial H_\mathrm{f}}{\partial k_x}\middle|u^{(\beta)}(\mathrm{k})\right\rangle \left\langle u^{(\beta)}(\mathrm{k})\middle|\frac{\partial H_\mathrm{f}}{\partial k_y}\middle|u^{(\alpha)}(\mathrm{k})\right\rangle\right]. \quad (23)$$

where $H_\mathrm{f} = H_\mathrm{f}(k,q,aA_0,b)$ in (1). The energy eigenvalues in (11) and the corresponding eigenstates are also dependent on the same set of parameters $(q,aA_0,b)$. In the zero-temperature limit, we obtain $\sigma_{AH} = C\left(\frac{e^2}{\hbar}\right)$ where $C = \int_{\mathrm{BZ}} \Omega_\alpha^z(k) \frac{d^2k}{(2\pi)^2}$ is the Chern number (an integer) - a topological invariant. The Chern number is a property of a material and is particularly important because, being an integer, it cannot be changed under continuous deformations of the system. It is also known as the Thouless-Kohmoto-Nightingale-Nijs (TKNN) number **[30,31]**. BC is the analogue of the magnetic field in momentum-space while the Berry connection $\mathrm{A}_\alpha(\boldsymbol{k})$ acts as a vector potential; that is, $\nabla_{\boldsymbol{k}} \times \mathrm{A}_\alpha(\boldsymbol{k}) = \Omega_\alpha(\mathbf{k})$. Therefore, the chern number C may also be expressed as $C = \sum_\alpha C_\alpha$, $C_\alpha = \frac{1}{2\pi}\oint d\boldsymbol{k}.\mathrm{A}_\alpha(\boldsymbol{k})$, where the closed loop corresponds to the encirclement of the Brillouin zone boundary. Upon borrowing the electromagnetic perspective, we can easily understand in the following simple way as to why $C$ should be an integer: Since $C$ basically is the momentum space flux integral, one may regard $C$ as the net number of Berry monopoles (whose charges do not cancel each other) within BZ. Furthermore, for the insulator system with spectral representation as in Figure 2(a) and 2(b) where the chemical potential lies within the band gap, because of the single-valued nature of the surface states above, corresponding change in the (Berry) phase factor after encircling the Brillouin zone boundary can only be an integer multiple of $2\pi$ or $2\pi$m, where m is an integer. This means $C$ can only take an integer value. Upon using (23) we will show below the same result calculating BC explicitly.

To simplify (23), we refer to the Heisenberg equation of motion $i\hbar\frac{d\hat{x}}{dt} = [\hat{x}, \hat{H}]$. In view of this equation, we find that the identity $\hbar\langle u^{(\alpha)}(k')|\hat{v}_j|u^{(\beta)}(\mathrm{k})\rangle = \left(\mathrm{E}_\alpha(\mathbf{k}') - \mathrm{E}_\beta(\mathbf{k})\right)\left\langle u^{(\alpha)}(k')\middle|\frac{\partial}{\partial k_j}\middle|u^{(\beta)}(\mathrm{k})\right\rangle$ is satisfied for a system in a periodic potential and its Bloch states as the eigenstates $|u^{(\alpha)}(\mathrm{k})\rangle$. Here the operator $\hbar^{-1}\frac{\partial H_\mathrm{f}}{\partial k_j} = \hat{v}_j$ represents the velocity in the $j = (x,y)$ direction. Upon using the identity above, the z-component of BC may be written in the form as

$$\Omega_{xy}(k) = \sum_\alpha\left(\frac{\partial A_{\alpha,y}}{\partial k_x} - \frac{\partial A_{\alpha,x}}{\partial k_y}\right) = -2\sum_\alpha Im\left\langle\frac{\partial u^{(\alpha)}(\mathrm{k})}{\partial k_x}\middle|\frac{\partial u^{(\alpha)}(\mathrm{k})}{\partial k_y}\right\rangle. \quad (24)$$

It is not difficult to see that for the present problem the imaginary part of $\left\langle\frac{\partial u^{(\alpha)}(\mathrm{k})}{\partial k_x}\middle|\frac{\partial u^{(\alpha)}(\mathrm{k})}{\partial k_y}\right\rangle$ is given by $Im(C_{x,1}C_{y,1}^{\dagger} - C_{x,2}C_{y,1}^{\dagger} - C_{x,1}C_{y,2}^{\dagger} + C_{x,2}C_{y,2}^{\dagger})$ where in view of Eq. (16)-(23) we obtain

$$C_{x,1} = N_\alpha^{-\frac{1}{2}}(\partial_x\Upsilon_1^{(\alpha)} - i\,\partial_x\Upsilon_2^{(\alpha)} \quad \partial_x\varkappa_1^{(\alpha)} - i\,\partial_x\varkappa_1^{(\alpha)} \quad \partial_x\,\Delta_1^{(\alpha)} - i\partial_x\,\Delta_2^{(\alpha)} \quad \partial_x\mathbb{Q}^{(\alpha)}), \tag{25}$$

$$C_{y,1}^{\dagger} = N_\alpha^{-\frac{1}{2}}(\partial_y\Upsilon_1^{(\alpha)} - i\,\partial_y\Upsilon_2^{(\alpha)} \quad \partial_y\varkappa_1^{(\alpha)} - i\,\partial_y\varkappa_2^{(\alpha)} \quad \partial_y\,\Delta_1^{(\alpha)} - i\partial_y\,\Delta_2^{(\alpha)} \quad \partial_y\mathbb{Q}^{(\alpha)})^{\dagger}, \tag{26}$$

$$C_{x,2} = 2^{-1}N_\alpha^{-\frac{3}{2}}\,\partial_x N_\alpha\,(\Upsilon_1^{(\alpha)} - i\,\Upsilon_2^{(\alpha)} \quad \varkappa_1^{(\alpha)} - i\,\varkappa_2^{(\alpha)} \quad \Delta_1^{(\alpha)} - i\,\Delta_2^{(\alpha)} \quad \mathbb{Q}^{(\alpha)}), \tag{27}$$

$$C_{y,2}^{\dagger} = 2^{-1}N_\alpha^{-\frac{3}{2}}\,\partial_y N_\alpha\,(\Upsilon_1^{(\alpha)} - i\,\Upsilon_2^{(\alpha)} \quad \varkappa_1^{(\alpha)} - i\,\varkappa_2^{(\alpha)} \quad \Delta_1^{(\alpha)} - i\,\Delta_2^{(\alpha)} \quad \mathbb{Q}^{(\alpha)})^{\dagger}. \tag{28}$$

The symbol $\partial_x$ ($\partial_y$) above stands for the differential coefficient $\frac{\partial}{\partial k_x}$ ( $\frac{\partial}{\partial k_y}$). As it is clear from (27) and (28) that ( $C_{x,2}C_{y,2}^{\dagger}$) is real, we need to find basically $Im(C_{x,1}C_{y,1}^{\dagger} - C_{x,2}C_{y,1}^{\dagger} - C_{x,1}C_{y,2}^{\dagger})$.

We show the outline of the calculation the *z*-component of the Berry curvature (BC) for the α^th occupied band $\Omega_{\mathrm{xy}}^{(\alpha)}(k) = 2\,Im(-C_{x,1}C_{y,1}^{\dagger} + C_{x,2}C_{y,1}^{\dagger} + C_{x,1}C_{y,2}^{\dagger})$ here. We obtain

$$-2Im\left(C_{x,1}C_{y,1}^{\dagger}\right) = 2N_\alpha^{-1}[-\partial_x\Upsilon_1^{(\alpha)}\partial_y\Upsilon_2^{(\alpha)} + \partial_x\Upsilon_2^{(\alpha)}\partial_y\Upsilon_1^{(\alpha)} - \partial_x\varkappa_1^{(\alpha)}\partial_y\varkappa_2^{(\alpha)} + \partial_x\varkappa_2^{(\alpha)}\partial_y\varkappa_1^{(\alpha)}$$
$$-\partial_x\Delta_1^{(\alpha)}\partial_y\Delta_2^{(\alpha)} + \partial_x\Delta_2^{(\alpha)}\partial_y\Delta_1^{(\alpha)}], \tag{29}$$

$$2\,Im(\,C_{x,2}C_{y,1}^{\dagger}) = N_\alpha^{-2}(\partial_x N_\alpha)\,[\Upsilon_1^{(\alpha)}\partial_y\Upsilon_2^{(\alpha)} - \Upsilon_2^{(\alpha)}\partial_y\Upsilon_1^{(\alpha)} + \varkappa_1^{(\alpha)}\partial_y\varkappa_2^{(\alpha)} - \varkappa_2^{(\alpha)}\partial_y\varkappa_1^{(\alpha)}$$
$$+\Delta_1^{(\alpha)}\partial_y\Delta_2^{(\alpha)} - \Delta_2^{(\alpha)}\partial_y\Delta_1^{(\alpha)}]\,, \tag{30}$$

$$2\,Im\left(C_{x,1}C_{y,2}^{\dagger}\right) = N_\alpha^{-2}(\partial_y N_\alpha)\,[\partial_x\Upsilon_1^{(\alpha)}\Upsilon_2^{(\alpha)} - \partial_x\Upsilon_2^{(\alpha)}\Upsilon_1^{(\alpha)} + \partial_x\varkappa_1^{(\alpha)}\varkappa_2^{(\alpha)} - \partial_x\varkappa_2^{(\alpha)}\varkappa_1^{(\alpha)}$$
$$+\partial_x\Delta_1^{(\alpha)}\Delta_2^{(\alpha)} - \partial_x\Delta_2^{(\alpha)}\Delta_1^{(\alpha)}], \tag{31}$$

In view of (29) - (31), we obtain

$$\Omega_{\mathrm{xy}}^{(\alpha)}(k) = N_\alpha^{-1}[\left(\partial_y\Upsilon_1^{(\alpha)} - N_\alpha^{-1}(\partial_y N_\alpha)\Upsilon_1^{(\alpha)}\right)\partial_x\Upsilon_2^{(\alpha)} - \left(\partial_y\Upsilon_2^{(\alpha)} - N_\alpha^{-1}(\partial_y N_\alpha)\Upsilon_2^{(\alpha)}\right)\partial_x\Upsilon_1^{(\alpha)}]$$
$$-N_\alpha^{-1}[\left(\partial_x\Upsilon_1^{(\alpha)} - N_\alpha^{-1}(\partial_x N_\alpha)\Upsilon_1^{(\alpha)}\right)\partial_y\Upsilon_2^{(\alpha)} - \left(\partial_x\Upsilon_2^{(\alpha)} - N_\alpha^{-1}(\partial_x N_\alpha)\Upsilon_2^{(\alpha)}\right)\partial_y\Upsilon_1^{(\alpha)}]$$
$$+N_\alpha^{-1}[\left(\partial_y\varkappa_1^{(\alpha)} - N_\alpha^{-1}(\partial_y N_\alpha)\varkappa_1^{(\alpha)}\right)\partial_x\varkappa_2^{(\alpha)} - \left(\partial_y\varkappa_2^{(\alpha)} - N_\alpha^{-1}(\partial_y N_\alpha)\varkappa_2^{(\alpha)}\right)\partial_x\varkappa_1^{(\alpha)}]$$
$$-N_\alpha^{-1}[\left(\partial_x\varkappa_1^{(\alpha)} - N_\alpha^{-1}(\partial_x N_\alpha)\varkappa_1^{(\alpha)}\right)\partial_y\varkappa_2^{(\alpha)} - \left(\partial_x\varkappa_2^{(\alpha)} - N_\alpha^{-1}(\partial_x N_\alpha)\varkappa_2^{(\alpha)}\right)\partial_y\varkappa_1^{(\alpha)}]$$

$$+ N_\alpha^{-1}[\left(\partial_y \Delta_1^{(\alpha)} - N_\alpha^{-1}(\partial_y N_\alpha)\Delta_1^{(\alpha)}\right)\partial_x \Delta_2^{(\alpha)} - \left(\partial_y \Delta_2^{(\alpha)} - N_\alpha^{-1}(\partial_y N_\alpha)\Delta_2^{(\alpha)}\right)\partial_x \Delta_1^{(\alpha)}]$$

$$-N_\alpha^{-1}[\left(\partial_x \Delta_1^{(\alpha)} - N_\alpha^{-1}(\partial_x N_\alpha)\Delta_1^{(\alpha)}\right)\partial_y \Delta_2^{(\alpha)} - \left(\partial_x \Delta_2^{(\alpha)} - N_\alpha^{-1}(\partial_x N_\alpha)\Delta_2^{(\alpha)}\right)\partial_y \Delta_1^{(\alpha)}]. \quad (32)$$

Now that we have calculated a formal expression for the BC of α band, what remains to be done is to calculate various derivatives in (32). This is a lengthy but straightforward task. In Figure 3 we have shown the contour plots of BC in the z-direction as a function of the wavenumber components ( $k_x, k_y$).The numerical values of the parameters used in the plots are $t_{d_1}$ = 1, $t_{f_1}$ = $-0.53$, $t_{d_2}$ = 0.01, $t_{f_2}$ = 0.01, $t_{d_3}$ = 0.001, $t_{f_3}$ = 0.001, $q$ = 0.1, $\epsilon_f = -$ 0.02, and μ = 0. The other values are as follow: $(a)$ $V = 0.6397,$ $b$ = 0.8426, and the amplitude of incident radiation $aA_0 = 1.3678.$ (b) $V = 0.6397,$ $b$ = 0.8426, and $aA_0 = 1.3678.$ $(c)$ V $= 0.3387,$ $b$ = 0.8132, and $aA_0 = 0.3500.$(d) $V = 0.6397,$ $b$ = 0.8123, and $aA_0 = 0.3500.$ The figures (a) and (c) correspond to left-handed circularly polarized light, while the figures (b) and (d) correspond to right-handed circularly polarized light. One may refer to Appendix C for an outline on how to obtain numerical value of '$b$' . Upon integrating BC on a k-mesh-grid of the surface Brillouin zone (BZ), we calculate the intrinsic anomalous Hall conductivity (AHC) $\sigma_{AH}$. This yields the Chern number ($C$). We find that AHC is $\sigma_{AH} = \mathrm{C}\left(\frac{e^2}{\hbar}\right)$ ( $(a)C = 0.9987, (b)C = 1.0028, (c)$ C $= 3.2918$ , and (d) C $= 2.3838$ ). Obviously enough, in (a) and (b), the Chern number has integer value ($C \approx 1$). However, we have non-integer values of the Chern number corresponding to Figures (c) and (d) as these two cases correspond to metallicity (see Figures 2(c) and 2(d)). We find that a higher value of the intensity of incident radiation indicates the possibility of the topological switching between the metallic state and the QAH state as the latter corresponds to the integer value of the Chern number $C$. The sign of the term $\mathrm{C}_3$ = $M$ in (1) does not seem to affect the value of $C$. That is, QAH effect is possible via periodically driven radiation in both the ferromagnetic (FM) and the anti-ferromagnetic (AFM) insulators. This is not quite surprising as there has been an attempt earlier to derive AFM Chern insulator from the Kane-Mele Hubbard model albeit in the non-centrosymmetric systems **[32].** Furthermore, recently it has been shown **[33]** that a monolayer CrO can be switched from an AFM Weyl semimetal to an AFM QAH insulator by applying strain. Also, in the case of the AFM material MoO monolayer the shear strain, upon breaking the time reversal symmetry, drives the system to have nontrivial electronic bands with the magnitude of the Chern number as unity **[34]**. Since the Neel temperature of the system is above the room temperature, it may be useful in spintronics applications. As for the other practical uses, the energy-efficient electronic devices **[35]** and the multichannel quantum computing **[36, 37]** could be facilitated by such topologically switchable materials.

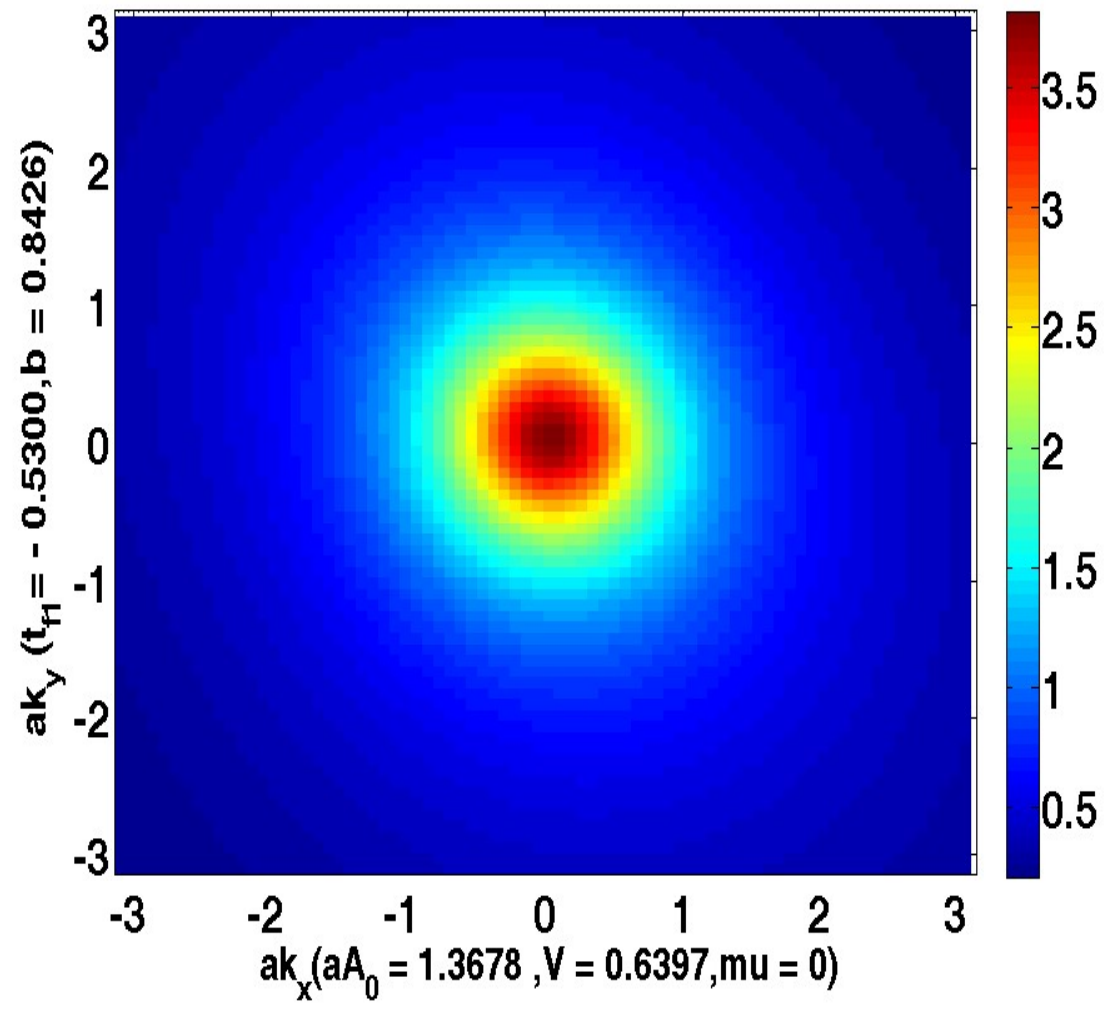

(a)$C = 0.9987$

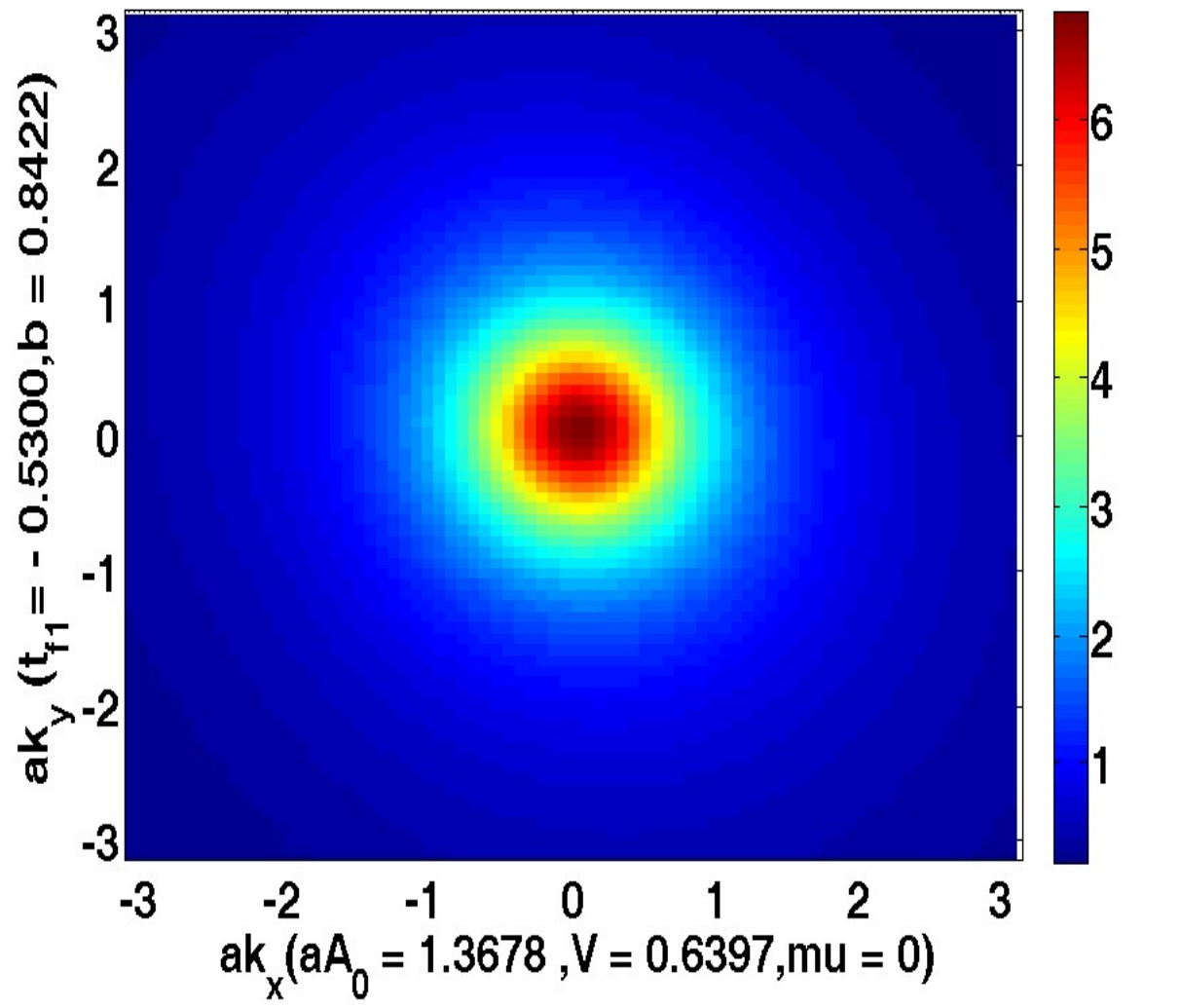

(b) $C = 1.0028$

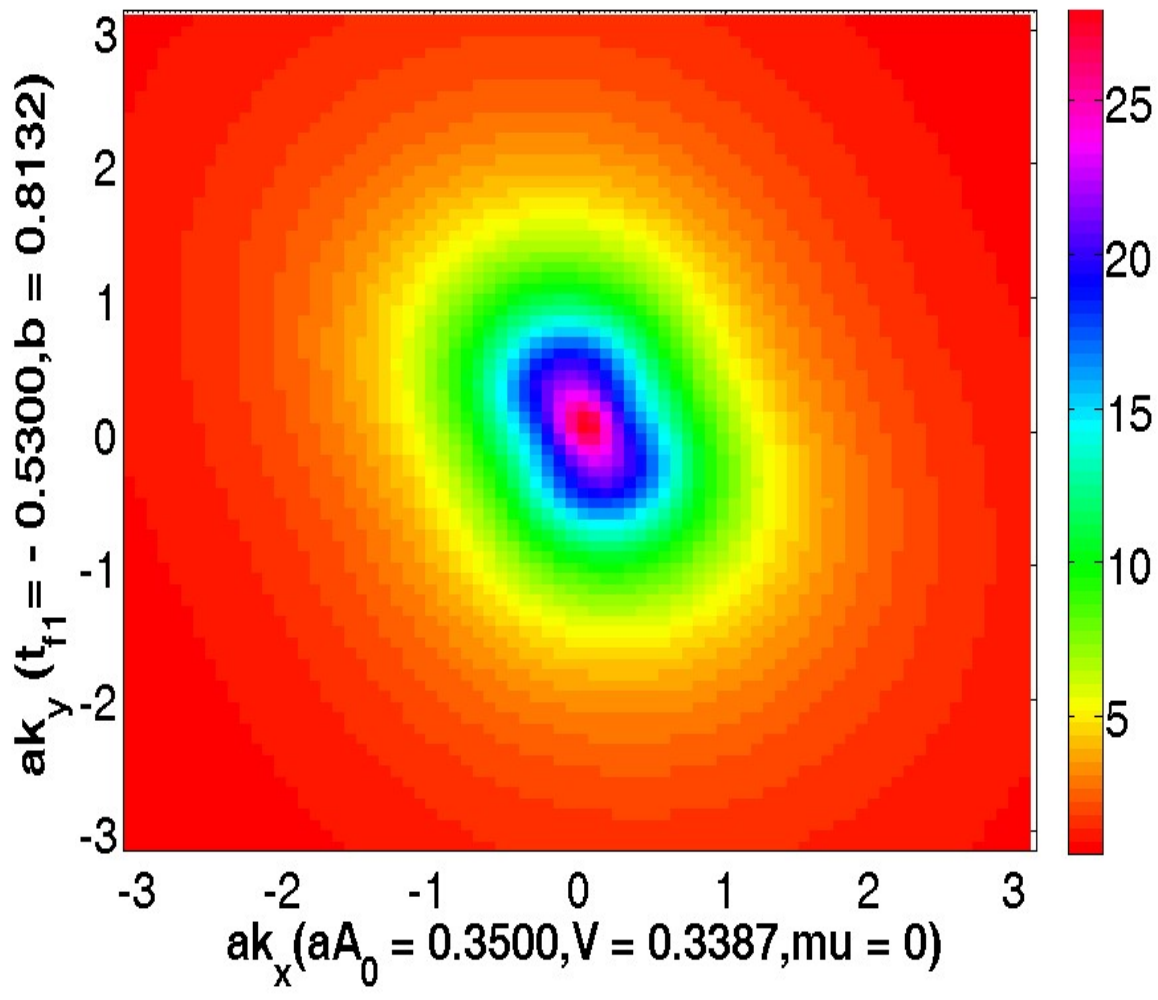

(c) $C = 3.2918$

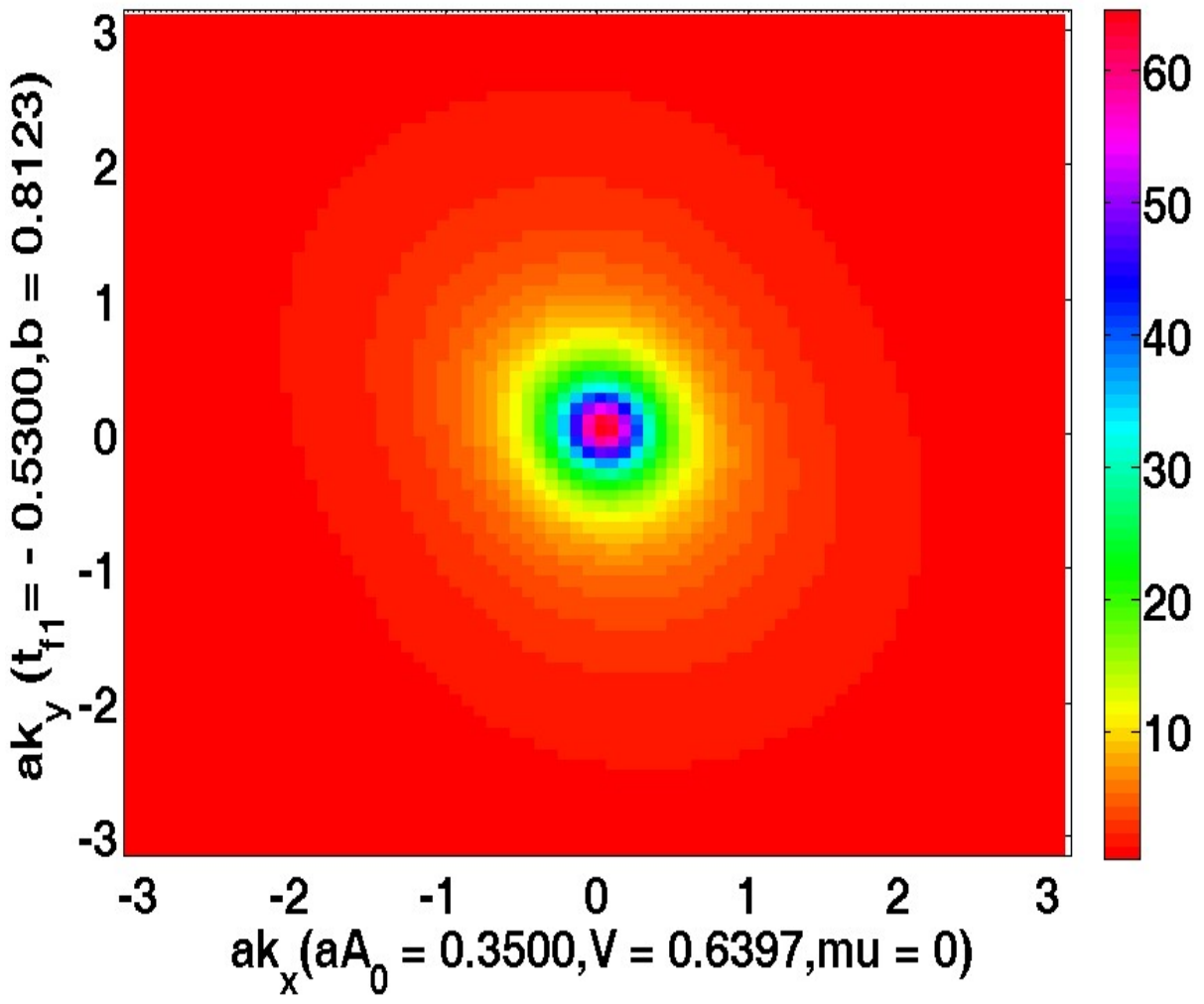

(d) $C = 2.3838$

**Figure 3.** The contour plots of the Berry curvature along z-direction as a function of the wavenumber components $(k_x, k_y)$.The numerical values of the parameters used in the plots are $t_{d_1} = 1$, $t_{f_1} = -0.53$, $t_{d_2} = 0.01$, $t_{f_2} = 0.01$, $t_{d_3} = 0.001$, $t_{f_3} = 0.001$, $q = 0.1$, $\epsilon_f = -0.02$, and $\mu = 0$. The other values are as follow: $(a)$and (b) $V = 0.6397$, $b = 0.8426$, and $aA_0 = 1.3678$. $(c)$ $V = 0.3387$, $b = 0.8132$, and $aA_0 = 0.3500$.(d) $V = 0.6397$, $b = 0.8123$, and $aA_0 = 0.3500$. The figures (a) and (c) correspond to left-handed circularly polarized light, while the figures (b) and (d) correspond to right-handed circularly polarized light. The values of the Chern number C are noted above.

## 5. Concluding remarks

The original time-dependent problem (with periodicity in external driving force) could be mapped into effective time-independent formulation in Floquet theory **[14-19]**. The theory was used extensively in the past in the theoretical studies of external driving on transport in various systems commissioning combination of the theory with Schrodinger equation **[38]**, Green's functions **[ 39-41]**, quantum master equation **[42,43]**, scattering matrix approach **[44]**, and so on. The combinations of the Floquet theory with dynamical mean field theory **[45]**, and slave boson protocol (SBP) **[46]** were also formulated for strongly correlated systems. Our approach in this paper is in acquiescence to the latter. In this article, we use a SB version of the Periodic Anderson Hamiltonian in combination with the Floquet theory to develop a microscopic approach to study the effects of external perturbations like electromagnetic fields on the non-trivial topological surface properties of $SmB_6$. The surface state, obtained by the evanescent wave approach (see Appendix A), is assumed to possess the comparable penetration depth like that in $Bi_2Se_3$. In the preceding sections we have shown that the periodic driving of circularly polarized light leads to a new avenue for obtaining novel QAH effect that may be inaccessible in static systems. For this purpose, the effective, static surface Hamiltonian (1) has been obtained using the high-frequency Floquet-Magnus expansion **[19, 47-49]**. These are the highlights of the article. It may be mentioned that in the high-frequency domain, other equivalent expansions such as van Vleck **[50,51]**, or Brillouin-Wigner **[52]** could also be used to describe the system surface in terms of a time-independent Hamiltonian.

As regards the future scope related to the system studied here, it may be noted that Michishita et al.**[53,54]** have reported recently that hybridization of *f* electrons with *d* electrons at low temperatures is essentially a non-Hermitian effect as the system undergoes crossover from localized *f* electrons state (at high temperatures) upon decreasing the temperature. They further reported that the crossover is accompanied by the appearance of exceptional points in the single-particle Greens function. The ubiquitous electron-electron, electron-impurity, and electron-phonon scatterings in an electronic system are responsible for the non-Hermicity of an open system resulting into certain degrees of gain and loss. These give rise to quasiparticles with finite lifetime. We are presently looking into this aspect of the problem of $SmB_6$ using an approach similar to that in refs. **[53,54]**. It is certainly quite interesting to analyze how does the incidence of the circularly polarized radiation affect the system in the presence of non-Hermiticity.

**<u>Statement Regarding Data Analyzed :</u>** All data used or analyzed during this study are included in this article.

---

## Appendix A

In order to obtain the surface state Hamiltonian , as the first step one makes an assumption, similar in spirit to that in ref. **[26],** that the plane surface $z = 0$ relates to the length and the breadth of the compound sample. As the second step of the evanescent wave approach, since $ak_z$ is not a good quantum number, one makes the replacement $ak_z \rightarrow -ia\,\partial_z$ in the various terms of the Hamiltonian $h(\boldsymbol{K} = (k_x, k, k_z), \mu, |b|)$ in section 2 and, by using the Taylor expansion, make them appear as polynomial operators $f(\partial_z)$. As the third step, one assumes that the states of the Hamiltonian are quasi-localized within the surface z = 0, and of the form $Aexp\left(-\frac{z}{d}\right)\ |u^{(\alpha)}(k)\rangle$ for $z>0$ and $Aexp\left(-\frac{|z|}{d}\right)\ |u^{(\alpha)}(k)\rangle$for $z<0$, where $k=\sqrt{k_x^2+k_y^2}$, $|u^{(\alpha)}(k)\rangle$ is eigenstate of $\alpha^{th}$ band of the surface Hamiltonian $H_f(k,q,aA_0,b)$ and $d$ is the surface-state penetration depth. The evanescent states are, thus, simply decaying for z > 0 and z < 0 if A is constant. The last step is to use the substitution rule $f(\partial_z)\, e^{qz}=f(q)e^{qz}$ where $q=d^{-1}$. For a value $d\sim$ 5-10 nm as in $Bi_2Se_3$, $\frac{a}{d}\sim 0.08-0.05$ (the lattice constant of $SmB_6$ is $a=0.413\ nm$)which ensures that the decaying term $exp\left(-\frac{z}{d}\right)$ and $exp\left(-\frac{|z|}{d}\right)\ \sim exp(-1)$for $|z|\sim$ 5-10 nm. We consider below this case.

We treat the model Hamiltonian matrix in (1) in the low-energy limit near $\bar{\Gamma}$ (0,0) point. In this limit, using Taylor expansion we write $\sin(ak_j)\ \rightarrow ak_j+O(a^3k_j^3), \cos(ak_j)\rightarrow (1-(\frac{1}{2}a^2k_j^2)+O(a^4k_j^4)), j=(x,y)$. Moreover, one may approximate $\sin(a/d)$ by $\left(\frac{a}{d}-\frac{(a/d)^3}{3!}+\cdots\right)$ and $\cos(a/d)\ by\ (1-\frac{(a/d)^2}{2}+\cdots)$. The low-energy surface Hamiltonian can now be conveniently written as $h_f=\frac{\epsilon(k,d,\mu,b)+\vartheta(k,b,d)}{2}(I^{4\times4}+\gamma^0)+\frac{\epsilon(k,d,\mu,b)-\vartheta(k,b,d)}{2}(I^{4\times4}-\gamma^0)+iA_1ak_x\gamma^2+A_1ak_y\gamma^0\gamma^2-iA_1(\frac{a}{d})\gamma^3$, where $A_1=\ 2Vb$. In the low-energy limit, the functions $\epsilon(k,\mu,d,b)=\frac{(\widetilde{E_k^d}(\mu,k)+\widetilde{E_k^f}(\mu,b,k))}{2}$ and $\vartheta(k,\mu,d,b)=\frac{(\widetilde{E_k^d}(\mu,k)-\widetilde{E_k^f}(\mu,b,k))}{2}$ have been approximated to $O(a^2k^2)$. The renormalized dispersions of *d*- and *f* -electrons, respectively, are given by the expressions $\widetilde{E_k^d}(\mu,k)=-\mu-\xi-\mathfrak{h}_d$ and $\widetilde{E_k^f}(\mu,b,k)=-\mu+\xi+\lambda-|b|^2\mathfrak{h}_f$, where $\mathfrak{h}_d=[2t_{d1}\,c_1(k)+4t_{d2}\,c_2(k)+8t_{d3}\,c_3(k)]$, $\mathfrak{h}_f=[2t_{f1}\,c_1(k)+4t_{f2}\,c_2(k)+8t_{f3}\,c_3(k)], c_1(k)=\sum_\mu\cos(ak_\mu)$, $c_2(k)=\sum_{\mu\neq\nu}(\cos k_\mu a\,.\cos k_\nu a)$, and $c_3(k)=\prod_\mu(\cos k_\mu a)$, $\mu=(x,y,z)$. The parameters $\{\lambda, \xi, |b|^2\}$ had been determined by the minimization of thermodynamic potential per unit volume in ref. **[22]** . Upon solving these simultaneous equations, we found that $\lambda=-6t_{f1}+6|b|^2t_{f1}$, $\xi=-3t_{d_1}+3t_{f_1}$.and $|b|^2\approx 0.80$, The outline of the fresh attempt to determine the crucial parameter $|b|^2$ ( $(a^2A_0^2)\neq 0$) will be given below in Appendix C. Upon using the same values of $\{\lambda, \xi\}$ as given above, in the first approximation, the functions $\epsilon(k,\mu,d,b)$ and $\vartheta(k,\mu,d,b)$ appearing in section 2 could be written as

$$\epsilon(k,d,\mu,b)=\epsilon_0(\mu,b)\ -D_1(b)a^2d^{-2}+\ D_2(b)a^2k^2+O(a^4d^{-4})+\ O(a^4k^4), \quad (A.1)$$

$$\vartheta(k,d,b)\ =\vartheta_0(b)-B_1(b)a^2d^{-2}\ +B_2(b)a^2k^2+O(a^4d^{-4})+\ O(a^4k^4), \quad (A.2)$$

$$\epsilon_0(\mu,b)=-\mu+[\frac{b^2}{2}\epsilon_f-3t_{d1}-3t_{f1}-6t_{d2}-6t_{f2}b^2-4t_{d3}-4t_{f3}\,b^2], \quad (A.3)$$

$$D_1(b)=D_2(b)=[\ \frac{t_{d1}+b^2t_{f1}}{2}+2(t_{d2}+b^2t_{f2})+2(t_{d3}+b^2t_{f3})], \quad (A.4)$$

$$\vartheta_0(b) = [-\frac{b^2\epsilon_f}{2} - 6t_{d2} + 6t_{f2}b^2 - 4t_{d3} + 4t_{f3}\, b^2], \quad (A.5)$$

$$B_1(b) = B_2(b) = [\frac{t_{d1}-b^2t_{f1}}{2} + 2(t_{d2} - b^2t_{f2}) + 2(t_{d3} - b^2t_{f3})], \quad (A.6)$$

in the low-energy limit, It may be mentioned in passing that in the case $d\sim$ 0.13 nm < $a$ (lattice constant) $\sim 0.4\, nm,$ we obtain $\frac{a}{d}$ $\sim 3.1416\ldots\ldots\sim\pi.$ Consequently, $e^{-|z|/d}$ will be nearly equal to $e^{-1}$ for very low value of $|z|$ ( ~ 0.1nm). We shall refer to this as the low penetration depth (LPD) case. We can present the Hamiltonian (1) in the basis $(d_{k,\uparrow}\ bc_{k,\sigma\downarrow}\ d_{k,\downarrow}\ bc_{k,\uparrow})^T$ in the non-block-diagonal form: $h(k, \mu, (aA_0), |b|) =$

$$\begin{pmatrix} \epsilon_{OP} + \vartheta_{OP}^{+} & A_{1OP}^{+}(ak_-) & 0 & -iA_{1O}^{+}\ (aq) \\ A_{1O}^{+}\ (ak_+) & \epsilon_{OP} - \vartheta_{OP}^{+} & -iA_{1O}^{-}\ (aq) & 0 \\ 0 & iA_{1OP}^{-}(aq) & \epsilon_{OP} + \vartheta_{OP}^{-} & -A_{1OP}^{-}(ak_-) \\ iA_{1OP}^{+}(aq) & 0 & -A_{1OP}^{-}(ak_+) & \epsilon_{OP} - \vartheta_{OP}^{-} \end{pmatrix} \quad (A.7)$$

Since in the LPD case the terms like $(iA_{1OP}^{\pm}(aq))$ will be absent, Eq. (A.7) assumes a block-diagonal form. The energy eigenvalues for this block-diagonal matrix are

$$\mathfrak{E}_{\mu}(s, k, b, I) = \epsilon_{OP}(k, b) \pm \sqrt{[\vartheta_{OP}^{(\mu)^2} + A_{1OP}^{(\mu)^2}((ak_x)^2 + (ak_y)^2)]} \quad (A.8)$$

where the index μ = +1(−1) for the left-handed (right-handed) circularly polarized radiation, respectively. The corresponding eigenvectors could be obtained from Eq. (16)-(23) dropping $(aq)$ involving terms. Upon using the formula $\Omega_{xy}(k) = = -2\sum_{\alpha} Im \left\langle \frac{\partial u^{(\alpha)}(k)}{\partial k_x} \middle| \frac{\partial u^{(\alpha)}(k)}{\partial k_y} \right\rangle$, it is not difficult to obtain the result $\Omega_{xy}(k) = \sum_{\alpha} \Omega_{xy}^{(\alpha)}(k)$ where

$$\Omega_{xy}^{(\alpha)}(k) = N_{\alpha}^{-1}\, ( A_{1OP}^{+\ 2} + A_{1OP}^{-\ 2})\, [\left(1 - N_{\alpha}^{-1}\partial_x N_{\alpha}(ak_x)\right) + \left(1 - N_{\alpha}^{-1}\partial_y N_{\alpha}(ak_y)\right)], \quad (A.9)$$

and $N_{\alpha} = [\, ( A_{1OP}^{+\ 2} + A_{1OP}^{-\ 2})\left((ak_x)^2 + (ak_y)^2\right) + (\mathfrak{E}_{\alpha} - \epsilon_{OP} - \vartheta_{OP}^{+})^2 + (\mathfrak{E}_{\alpha} - \epsilon_{OP} - \vartheta_{OP}^{-})^2].$

**Appendix B**

Analogous to the Bloch theory involving quasi-momentum, a solution/wave function $|\psi\ (t)\rangle = e^{-i\acute{\varepsilon}t}|\ \psi_f\ (t)\rangle$ involving the Floquet quasi-energy $\acute{\varepsilon}$ could be written down for the time-dependent

Schrodinger equation of the system, where the Floquet state satisfies $|\psi_f(t)\rangle = |\psi_f(t+T)\rangle$. The periodicity implies that $|\psi_f(t)\rangle$ could be expanded in a Fourier series: $|\psi_f(t)\rangle = \sum_r \exp(-ir\omega t)\,|\psi_f^r\rangle$ where r is an integer. Then the wave function $|\psi(t)\rangle$, in terms of the quasi-energy $\acute{\varepsilon}$, has the form $|\psi(t)\rangle = \sum_r \exp\left(-i\left(\frac{\acute{\varepsilon}}{\hbar} + r\omega\right)t\right|\psi_f^r\rangle$. This makes us arrive at an infinite dimensional eigenvalue equation in the Sambe space (the extended Hilbert space)**[26]**: $\sum_r h_{f,r,s}\,|\psi_{f,n}^s\rangle = \acute{\varepsilon}_n|\psi_{f,n}^s\rangle$.The matrix element $h_{f,\alpha,\beta}$ is given by $h_{f,\alpha,\beta} = \alpha\hbar\omega\delta_{\alpha,\beta} + \frac{1}{T}\int_0^T h_f(t)e^{i(\alpha-\beta)\omega t}dt$, where $(\alpha,\beta)$ are integers. This is the Floquet surface state Hamiltonian matrix element. In view of the Floquet theory **[14-19]**, we can now write a static effective Hamiltonian $H_{\mathrm{f}}^{Floquet}(k,d,aA_0,\mu,b)$, in the off-resonant regime using the Floquet-Magnus expansion **[17]:** $H_{\mathrm{f}}^{Floquet}(k,d,aA_0,\mu,b) = h_{f,0,0} + \frac{[h_{f,0,-1},h_{f,0,1}]}{\hbar\omega} + O(\omega^{-2})$,where $h_{f,n,m} = \frac{1}{T}\int_0^T h_f(t)e^{i(n-m)\omega t}dt$ with $n \neq m$. In the low-energy and high frequency limit, the terms $h_{f,0,0}$ is given by $h_{f,0,0} = h_{\mathrm{f}}(k,q,\mu,b) + D_1(b)(a^2A_0^2)I_{4\times4} + B_1(b)\ (a^2A_0^2)\,\sigma_{12}$. The remaining two terms $h_{f,0,1}$, and $h_{f,0,-1}$ are polarization handedness dependent, e.g. $h_{f,0,1} = iD_1(b)\chi I^{4\times4}$ $+iB_1(b)\chi\sigma_{12} + (i/2)\,A_1(aA_0)\sigma_{23} + (i/2)\,A_1(aA_0)\,e^{i\psi}\sigma_{31}$, $\chi = a^2\,(k_x + e^{i\psi}k_y)A_0$ . The term $h_{f,0,-1}$ could be obtained replacing $i$ by $(-i)$ in $h_{f,0,1}$. Here $\sigma_{\nu\rho} = \left(\frac{i}{2}\right)[\gamma_\nu, \gamma_\rho]$, and the covariant form $\gamma_\mu = \eta_{\mu\nu}\gamma^\nu = (\gamma^0, -\gamma^1, -\gamma^2, -\gamma^3, \gamma^5)$. These commutators are expressible in terms of the tensor product of Pauli matrices in the orbital-spin basis. While $\boldsymbol{\sigma} = (\sigma_x, \sigma_y, \sigma_z)$ acts on the real spin, $\boldsymbol{\tau} = (\tau_x, \tau_y, \tau_z)$ acts on the orbital degree of freedom. The matrices $\boldsymbol{\tau}$ and $\boldsymbol{\sigma}$ are acting together in the space of bands yielding spectral gaps in Figure 1.

We can write the (anti-unitary) time-reversal operator $\Theta = UK$ where $U$ is a unitary operator. Furthermore, for a spin-1/2 particle, flipping the spin coincides with the time-reversal. This means $\Theta\,\widehat{\boldsymbol{S}} = -\widehat{\boldsymbol{S}}$ where $\widehat{\boldsymbol{S}} = \frac{1}{2}\,\widehat{\boldsymbol{\sigma}}$ and $\widehat{\boldsymbol{\sigma}}$ is the vector of Pauli matrices. In view of these, one may choose $\Theta = i\tau_0 \otimes \sigma_y K$, where $\tau_0$ is the identity matrix. The operator $K$ stands for the complex conjugation. Upon making use of the results $\Theta\,\widehat{A}\,\Theta^{-1} = \widehat{A}$, $\Theta\,\widehat{B}\,\Theta^{-1} = -\widehat{B}$ ,and so on, where $\widehat{A} = \tau_0\otimes\sigma_0,\ \tau_z\otimes\sigma_y, \ldots\ldots$ and $\widehat{B} = \tau_0\otimes\sigma_y, \ldots$, we find that $\Theta\,H_{\mathrm{f}}^{Floquet}(ak_x, ak_y, b, aA_0)\Theta^{-1}$ $= H_{\mathrm{f}}^{Floquet}(-ak_x, -ak_y, b)$ only when $\psi = 0$ or $\pi$, that is, when the radiation field is linearly polarized . Thus, for the linearly polarized radiation, the time reversal symmetry (TRS) is not broken. However, when $\psi = \frac{\pi}{2}$ $or$ $-\frac{\pi}{2}$, that is the radiation is circularly polarized, TRS is broken. The reason being in these cases

$$\Theta\,H_{\mathrm{f}}^{Floquet}(ak_x, ak_y, b, aA_0)\Theta^{-1} = H_{\mathrm{f}}^{Floquet}(-ak_x, -ak_y, b) + (4a^2A_0^2 sin\psi/\hbar\omega)\times$$

$$\{A_1 ak_x\,\tau_0\otimes\sigma_{\mathrm{x}} + A_1 ak_y\,\tau_0\otimes\sigma_{\mathrm{y}}\}\,D_1(b) + \left(\frac{4a^2A_0^2\,A_1^2 sin\psi}{\hbar\omega}\right)\tau_0\otimes\sigma_{\mathrm{z}}, \tag{B.1}$$

We have adopted the same basis as in the appendix A to write (B.1). As shown above, due to the broken TR symmetry, this system can be regarded as a Chern insulator- a TR symmetry broken topological insulator with a non-zero Chern number. It is worthwhile to note here that in the absence of spin-orbit interaction the Anderson model is both time reversal and inversion symmetric.

**Appendix C**

As noted in ref. **[23]**, in the SB-framework based extension of the mean-field-theoretic version of PAM **[3,17,18]** used here, one makes the replacement $f_\alpha(\boldsymbol{i}) \to c_\alpha(\boldsymbol{j}) b^\dagger(\boldsymbol{j})$ where $f_\alpha(\boldsymbol{j})$ is annihilation operator for an *f*-electron, and $\alpha$ ($\alpha = \uparrow, \downarrow$) is a pseudospin quantum number corresponding to the two possible values of the projection of total angular momentum in the lowest-lying is $|\Gamma_8^{f(2)}\rangle = |\alpha = \pm\frac{1}{2}\rangle$ (in the case of $SmB_6$). Here $c_\alpha(\boldsymbol{j})$ is a pseudo-spinful fermion annihilation operator, and $b^\dagger(\boldsymbol{j})$ is a spinless slave-boson creation operator at a site $\boldsymbol{j}$. This gives rise to extension in the Hilbert space of the system. The complications associated with the large on-site repulsion between the *f*-electrons, implying no double occupancy of a site, is conveniently circumvented by imposing the holonomic constraint $[\sum_\alpha \langle c_\alpha^\dagger(\boldsymbol{j}) c_\alpha(\boldsymbol{j})\rangle + \langle b^\dagger(\boldsymbol{j}) b(\boldsymbol{j})\rangle] = 1$. This holonomic constraint is then imposed in a mean-field fashion using a Lagrange multiplier λ. For the formation of Kondo singlet states, one requires the number of *f*-fermions equal to the number of *d*- fermions which is enforced by the auxiliary chemical potential $\xi$; the chemical potential $\mu$ of the fermion number is a free parameter. Upon assuming no spatial dependence of the boson operators and replacing them by their expectation value ($\langle b^\dagger(\boldsymbol{j}) b(\boldsymbol{j})\rangle \to |b|^2$) for simplicity, we find that the model Hamiltonian involves the set of three unknown (non-free) parameters $\{\lambda, \xi, |b|^2\}$. Here $b$ may be complex as the density distribution of the slave boson-condensate is represented by a wavefunction with a well-defined amplitude and phase. The anticommutation relation $\{f_\alpha, f_\beta^\dagger\} = \delta_{\alpha\beta}$ implies that $|b|^2\{c_\alpha, c_\beta^\dagger\} = \delta_{\alpha\beta}$. As mentioned above, the parameters $\{\lambda, \xi, |b|^2\}$ have been determined by the minimization of thermodynamic potential per unit volume $\Omega_{sb} = -(\beta V)^{-1} \ln Tr \exp(-\beta H_{sl.boson}(|b|^2, \lambda, \xi))$: $\partial\Omega_{sb}/\partial|b| = 0$, $\partial\Omega_{sb}/\partial\lambda = 0$, and $\partial\Omega_{sb}/\partial\xi = 0$ in ref. **[22]** . Here $\beta$ denotes the inverse of the product of temperature $T$ and Boltzmann constant $k_B$. Upon solving these simultaneous equations, it was found that $\lambda = -6t_{f1} + 6|b|^2 t_{f_1}$, $\xi = -3t + 3t_{f1}$, where $t = t_{d1}$, and $t_{f1} < 0$ ($t_{f1} > 0$) for the insulating ( conducting) bulk. The estimated numerical value of the parameter $|b|$ for the bulk was found to be $|b|^2 \sim 0.80$. The numerical value of $|b|^2$ depends on the choice of the values of the hopping parameters (and the hybridization parameter). In the present case, since $(a^2 A_0^2) \neq 0$ , $|b|^2$ changes with electric field (the electric field amplitude $\boldsymbol{E_0} = \boldsymbol{A_0}\omega$ (see third paragraph in section 2)). In view of this, we find it necessary to recalculate $|b|^2$. In what follows, we focus on the equation $\partial\Omega_{sb}/\partial|b| = 0$ for this purpose. We write the expectation value of the slave-boson mean-field Hamiltonian $h(k, \mu, (aA_0), |b|)$ in (A.7) as

$$\langle h(k, \mu, (aA_0), |b|)\rangle = \sum_k (\epsilon_{OP} + \vartheta_{OP}^+) \langle d_{k\uparrow}^\dagger d_{k,\uparrow}\rangle + (\epsilon_{OP} + \vartheta_{OP}^-)\langle d_{k\downarrow}^\dagger d_{k,\downarrow}\rangle$$

$$+|b|^2 (\epsilon_{OP} - \vartheta_{OP}^+)\langle c_{k\downarrow}^\dagger c_{k,\downarrow}\rangle + |b|^2(\epsilon_{OP} - \vartheta_{OP}^-)\langle c_{k\uparrow}^\dagger c_{k,\uparrow}\rangle + \lambda N(|b|^2 - 1) + \langle Y_{k,\zeta=\uparrow,\downarrow}\rangle, \quad \text{(C.1)}$$

$$\langle Y_{k,\zeta=\uparrow,\downarrow}\rangle = |b|\, A_{1OP}^+(ak_-)\langle d_{k\uparrow}^\dagger c_{k,\downarrow}\rangle + |b|(-iA_{1OP}^+(aq))\langle d_{k\uparrow}^\dagger c_{k,\uparrow}\rangle + |b|\, A_{1OP}^+(ak_+)\langle c_{k\downarrow}^\dagger d_{k,\uparrow}\rangle$$

$$+|b|(-iA_{1O}^{-}\ (aq))\langle c_{k.\downarrow}^{\dagger}\, d_{k,\downarrow}\rangle + |b|(iA_{1OP}^{-}(aq))\langle d_{k.\downarrow}^{\dagger}\, c_{k,\downarrow}\rangle$$

$$+|b|\left(-A_{1OP}^{-}(ak_{-})\right)\langle d_{k.\downarrow}^{\dagger}\, c_{k,\uparrow}\rangle + |b|(iA_{1OP}^{+}(aq))\langle c_{k.\uparrow}^{\dagger}\, d_{k,\uparrow}\rangle + |b|\,(-A_{1OP}^{-}(ak_{+}))\,\langle c_{k.\downarrow}^{\dagger}\, d_{k,\downarrow}\rangle, \quad \text{(C.2)}$$

where $N$ is the number of fermions. The various functions in (C.1) and (C.2) are given by Eqs. (2)-(4) and (A.1) to (A.6). Upon replacing $|b|$ by another dimensionless quantity $b_1 \equiv \frac{|b|V}{t}$ for our convenience and using values of ($\lambda$, ξ) above in the first approximation, the equation $\frac{\partial \Omega_{sb}}{\partial |b|} = 0$ appears as

$$2\left(-6t_{f1} + 6{b_1}^2\left(\frac{t^2}{V^2}\right)t_{f1}\right)\left(\frac{b_1 t}{V}\right) = N^{-1}\sum_{k,\zeta=\uparrow,\downarrow}\left\{\left(\frac{V}{t}\right)\frac{\partial(\epsilon_{OP}(b_1)\langle d_{k.\zeta}^{\dagger}\, d_{k,\zeta}\rangle)}{\partial b_1} + \frac{\partial({b_1}^2 t V^{-1}\epsilon_{OP}(b_1)\langle c_{k.\zeta}^{\dagger}\, c_{k,\zeta}\rangle)}{\partial b_1}\right\} + N^{-1}\sum_k\left\{\left(\frac{V}{t}\right)\frac{\partial(\vartheta_{OP}^{+}(b_1)\,\langle d_{k.\uparrow}^{\dagger}\, d_{k,\uparrow}\rangle) + \vartheta_{OP}^{-}(b_1)\langle d_{k.\downarrow}^{\dagger}\, d_{k,\downarrow}\rangle)}{\partial b_1}\right\}$$

$$-N^{-1}\sum_k\left\{\frac{{b_1}^2 t V^{-1}\partial\langle\vartheta_{OP}^{-}\,\langle c_{k.\uparrow}^{\dagger}\, c_{k,\uparrow}\rangle\rangle + {b_1}^2 t V^{-1}\,\vartheta_{OP}^{+}\langle c_{k.\downarrow}^{\dagger}\, c_{k,\downarrow}\rangle)}{\partial b_1}\right\} + N^{-1}\sum_k\left\{\left(\frac{V}{t}\right)\frac{\partial(\langle Y_{k,\zeta=\uparrow,\downarrow}(b_1)\rangle)}{\partial b_1}\right\}. \quad \text{(C.3)}$$

in the zero-temperature and high frequency limit, mentioned in Appendix B. Since $\left(\frac{V}{t}\right)$ is a free parameter, it is convenient to write the equation $\frac{\partial \Omega_{sb}}{\partial |b|} = 0$ in terms of $b_1$ easily. The derivatives of the functions in (C.3) are given by $\frac{\partial \epsilon_0(b_1)}{\partial b_1} = 2\left(\frac{t}{V}\right)^2 b_1(\epsilon_f - 6t_{f2} - 4t_{f3})$, $\frac{\partial D_1(b)}{\partial |b|} = \left(\frac{t}{V}\right)^2 b_1\ (\ t_{f1} + 4t_{f2} + 4t_{f3})$, and so on. Finding a solution for $b_1$ solving Eq.(C.3), however, is no mean feat as it involves, apart from these derivatives, the expectation values $\langle d_{k.\zeta}^{\dagger}\, d_{k,\zeta}\rangle$, $\langle\, c_{k.\zeta}^{\dagger}\ c_{k,\zeta}\rangle$, etc. which are also $b_1$ dependent. These values are calculated within the finite-temperature formalism in a manner similar to the slave bosonic framework reported in refs. **[22]** and **[25]**. We give the outline of the calculation below: Since the Hamiltonian in (C.1) is completely diagonal one can easily write down the equations of motion for the operators$\{d_{k.\zeta}(\tau), c_{k,\zeta}(\tau)\}$,where the time evolution of an operator $\hat{O}$ is given by $\hat{O}(\tau)$=exp($h(k,\mu,(aA_0),|b|)\tau$) $\hat{O}$ exp($-h(k,\mu,(aA_0),|b|)\tau$), as the first step. We obtain

$$\frac{\partial d_{k,\uparrow}(\tau)}{\partial \tau} = (\epsilon_{OP} + \vartheta_{OP}^{+})d_{k,\uparrow}(\tau) + \left(\frac{t}{V}\right)b_1\, A_{1OP}^{+}(ak_{-})c_{k,\downarrow}(\tau) + \left(\frac{t}{V}\right)b_1(-iA_{1OP}^{+}(aq))c_{k,\uparrow}(\tau),$$

$$\frac{\partial d_{k,\downarrow}(\tau)}{\partial \tau} = (\epsilon_{OP} + \vartheta_{OP}^{-})d_{k,\downarrow}(\tau) + \left(\frac{t}{V}\right)b_1\left(-A_{1OP}^{-}(ak_{-})\right)c_{k,\uparrow}(\tau) + \left(\frac{t}{V}\right)b_1(iA_{1OP}^{-}(aq))c_{k,\downarrow}(\tau),$$

$$\frac{\partial\left(c_{k,\uparrow}(\tau)\right)}{\partial \tau} = (\epsilon_{OP} - \vartheta_{OP}^{-})\,c_{k,\uparrow}(\tau) + (-A_{1OP}^{-}(ak_{+}))\,d_{k,\downarrow}(\tau) + (iA_{1OP}^{+}(aq))d_{k,\uparrow}(\tau),$$

$$\frac{\partial\left(c_{k,\downarrow}(\tau)\right)}{\partial \tau} = (\epsilon_{OP} - \vartheta_{OP}^{+})c_{k,\downarrow}(\tau) + A_{1OP}^{+}(ak_{+})d_{k,\uparrow}(\tau) + (-iA_{1OP}^{-}(aq))d_{k,\downarrow}(\tau). \quad \text{(C.4)}$$

As the second step, Eq. (C.4) is used to obtain the equation of motion of the temperature functions $G(k\zeta, k\zeta,\tau) = -\langle T_\tau\{d_{k.\zeta}(\tau)d^{\dagger}_{k,\zeta}(0)\rangle$, $F(k\zeta, k\zeta,\tau) = -\langle T_\tau\{c_{k,\zeta}(\tau)d^{\dagger}_{k,\zeta}(0)\rangle$, etc. where $T_\tau$ is the time-ordering operator which arranges other operators from right to left in the ascending order of imaginary time $\tau$. We obtain, for example,

$$\frac{\partial G(\mathrm{k}\uparrow,\mathrm{k}\uparrow,\tau)}{\partial\tau} = (\epsilon_{OP} + \vartheta_{OP}^{+})G(\mathrm{k}\uparrow,\mathrm{k}\uparrow,\tau) + (\frac{tb_1}{V})\, A_{1O}^{+}\ (ak_-)\, F(k\downarrow, k\uparrow,\tau) + (\frac{tb_1}{V})\,(-iA_{1O}^{+}\ (aq))$$

$$\times\ F(k\uparrow, k\uparrow,\tau) - \delta(\tau),$$

$$\frac{\partial F(\mathrm{k}\uparrow,\mathrm{k}\uparrow,\tau)}{\partial\tau} = (\epsilon_{OP} - \vartheta_{OP}^{-})F(\mathrm{k}\uparrow,\mathrm{k}\uparrow,\tau) + (-(\frac{tb_1}{V})\, A_{1OP}^{-}(ak_+))\ G(k\downarrow, k\uparrow,\tau) + (\frac{tb_1}{V})\,(iA_{1O}^{+}\ (aq))$$

$$\times\ G(k\uparrow, k\uparrow,\tau),$$

$$\frac{\partial F(\mathrm{k}\downarrow,\mathrm{k}\uparrow,\tau)}{\partial\tau} = (\epsilon_{OP} - \vartheta_{OP}^{+})F(\mathrm{k}\downarrow,\mathrm{k}\uparrow,\tau) + ((\frac{tb_1}{V})\, A_{1O}^{+}\ (ak_+))\ G(\mathrm{k}\uparrow,\mathrm{k}\uparrow,\tau) + (\frac{tb_1}{V})\,(-iA_{1OP}^{-}(aq\ ))$$

$$\times\ G(k\downarrow, k\uparrow,\tau),$$

$$\frac{\partial G(\mathrm{k}\downarrow\ ,\mathrm{k}\uparrow,\tau)}{\partial\tau} = (\epsilon_{OP} + \vartheta_{OP}^{-})G(\mathrm{k}\ \downarrow\ ,\mathrm{k}\uparrow,\tau) + \left(-(\frac{tb_1}{V})A_{1OP}^{-}(ak_-)\right) F(\mathrm{k}\uparrow,\mathrm{k}\uparrow,\tau) + (\frac{tb_1}{V})(iA_{1OP}^{-}(aq))$$

$$\times F(\mathrm{k}\downarrow,\mathrm{k}\uparrow,\tau). \quad \text{(C.5)}$$

The third step is to write down equations like (C.5) in terms of the Matsubara propagators (MPs), viz. $G(k\zeta, k\zeta, z = i\omega_n)$ and $F(k\zeta, k\zeta, z = i\omega_n)$ where $\omega_n = \frac{(2n+1)\pi}{\beta}$ and n is an integer. The equations for MPs are algebraic equations which can be solved easily. One obtains, for example,

$$G(\mathrm{k}\uparrow,\mathrm{k}\uparrow, z) = \frac{f(k,z=z_1)}{z-z_1} + \frac{f(k,z=z_2)}{z-z_2} + \frac{f(k,z=z_3)}{z-z_3} + \frac{f(k,z=z_4)}{z-z_4} \quad \text{(C.6)}$$

$$f(z) = ((z+\epsilon_{OP})^2 - {\vartheta_{OP}^{-}}^2)(\ z+\epsilon_{OP} - \vartheta_{OP}^{+}) - (\ z+\epsilon_{OP} - \vartheta_{OP}^{-})[\ (\frac{tb_1}{V})\,(A_{1OP}^{-}(aq\ ))]^2$$

$$-(\ z+\epsilon_{OP} - \vartheta_{OP}^{+})[\ (\frac{tb_1}{V})\,(A_{1OP}^{-}(ak\ ))]^2. \quad \text{(C.7)}$$

In the low-energy and high frequency limit, the poles of the propagator in (C.6) are $z_1, z_3 \approx -\epsilon_{OP} \pm \Xi_+;\ z_2, z_4 \approx -\epsilon_{OP} \pm \Xi_-$ , where $\Xi_{+,-} = {\vartheta_{OP}^{\pm}}^2 + (\frac{1}{2})(\ {A_{1OP}^{+}}^2 + {A_{1OP}^{-}}^2)(\frac{tb_1}{V})^2\ \ (\ (aq)^2 + (ak)^2)$ . Once all such equations (for the Fourier transforms) are solved, the last step is to obtain the expectation values. For example,

$$\langle d_{\mathrm{k}.\uparrow}^{\dagger}\, d_{\mathrm{k},\uparrow}\rangle\ = -\langle T_\tau\{d_{\mathrm{k}.\uparrow}\, d^{\dagger}{}_{\mathrm{k},\uparrow}(0^+)\rangle = \frac{1}{\beta}\ \ \sum_n \exp(z\eta)\, G(k\uparrow, k\uparrow, z = i\omega_n), \eta =\ 0^+. \quad \text{(C.8)}$$

This approach ensures that the thermal averages in the Eq. (C.5) above are determined in a self-consistent manner. In the low-temperature limit, the summations in (C.3) appear as $N^{-1}\sum_{ak<a\ _F}\langle\dots\dots\rangle$ , where the multiplication of the Fermi momentum $k_F$ with the lattice constant $'a'$ yields a dimensionless quantity. We assume the value of the Fermi velocity $(v_F)\sim 0.5\times 10^3\ \frac{m}{s}$ **[25]**, and the effective mass of the electron $m^*\sim 200\ m_e$**[55]**, where $m_e$ is the bare mass of an electron. This leads to the value of the Fermi wave vector $ak_F \sim 0.4$. The summation/integration $N^{-1} \sum_{ak<a_F} \langle \dots \dots \rangle$ becomes somewhat manageable with these experimental inputs. We solve Eq. (C.3) numerically. For this purpose, we assume the numerical values of the parameters in (C.3) as $t_{d_1} = 1$, $t_{f_1} = -0.53$, $t_{d_2} = 0.01$, $t_{f_2} = 0.01$, $t_{d_3} = 0.001$, $t_{f_3} = 0.001$, $q = 0.1$, $\epsilon_f = -0.02$, and $\mu = 0$. The other values are as follow: $(i)$ For $aA_0 = 1.3678$, and $V = 0.6397$, we obtain $b = 0.8426$. $(ii)$For $aA_0 = 0.3500$, and $V = 0.3387$, we obtain $b = 0.8132$.$(iii)$ For $aA_0 = 0.3500$, and $V = 0.6397$, we obtain $b = 0.8123$. Since, the estimated numerical value of this parameter for the bulk was found to be $|b|^2 \sim 0.80$ earlier **[22],** the value(s) obtained here suggests that the parameter is mildly affected by the intensity of incident radiation. As $|b| \longrightarrow 0$ implies that the system is a non-interacting lattice gas mixture of itinerant *d*- and non-hopping *f*- fermions with no topological dispensation and $|b| \longrightarrow 1$ implies no correlation effect on the hopping parameter $t_{f_1}$ and the hybridization parameter $V$, the obvious conclusion is correla tion effects on ( $t_{f_1}, V$) are slightly stronger on the surface involving CPRF than in the bulk as in ref.**[22]**.